\documentclass[conference]{IEEEtran}
\IEEEoverridecommandlockouts
\usepackage{cite}
\usepackage{amsmath,amssymb,amsfonts}
\usepackage{graphicx}
\usepackage{textcomp}
\usepackage{xcolor}
\def\BibTeX{{\rm B\kern-.05em{\sc i\kern-.025em b}\kern-.08em
    T\kern-.1667em\lower.7ex\hbox{E}\kern-.125emX}}

\usepackage[ruled, lined, linesnumbered, commentsnumbered, longend]{algorithm2e}
\usepackage{algpseudocode}

\newcommand{\VEC}[1]{\boldsymbol{#1}}
\newcommand{\MAT}[1]{\boldsymbol{#1}}

\newcommand{\PR}[1]{\left ( {#1} \right ) }
\newcommand{\BK}[1]{\left [ {#1} \right ] }
\newcommand{\CR}[1]{\left \{ {#1} \right \} }

\newcommand{\ABS}[1]{\left | {#1} \right |}

\newcommand{\EXPECTATION}[2]{\mathbb{E}_{#1} \BK{#2}}
\newcommand{\NORM}[1]{\left\lVert#1\right\rVert}

\newcommand{\PDERIV}[2]{ \frac{\partial \, {#1} }{\partial \, {#2}} }
\newcommand{\PDERIVFLAT}[2]{ {\partial \, {#1} }/{\partial \, {#2}} }
\newcommand{\DIAG}[1]{\textbf{\textrm{diag}} \BK{#1} }

\newcommand{\TRACE}[1]{\textbf{\textrm{tr}}\BK{#1} }

\DeclareMathOperator*{\argmax}{arg\,max}
\DeclareMathOperator*{\argmin}{arg\,min}

\newcommand{\plus}{\raisebox{0\height}{\scalebox{.55}{+}}}
\newcommand{\minus}{\raisebox{0\height}{\scalebox{.85}{-}}}

\newcommand{\pluscap}{\texttt{+}}
\newcommand{\minuscap}{\texttt{-}}

\SetKwInput{KwRequire}{Require}

\renewcommand{\IEEEbibitemsep}{0pt plus 0.5pt}
\makeatletter
\IEEEtriggercmd{\reset@font\normalfont\fontsize{7.9pt}{8.40pt}\selectfont}
\makeatother
\IEEEtriggeratref{1}

\begin{document}

\title{Constant Directivity Loudspeaker Beamforming\\
}

\author{\IEEEauthorblockN{Yuancheng Luo}
\IEEEauthorblockA{\textit{Amazon Inc.}\\
luoyuancheng@gmail.com
}
}

\maketitle

\begin{abstract}
Loudspeaker array beamforming is a common signal processing technique for acoustic directivity control and robust audio reproduction. Unlike their microphone counterpart, loudspeaker constraints are often heterogeneous due to arrayed transducers with varying operating ranges in frequency, acoustic-electrical sensitivity, efficiency, and directivity. This work proposes a frequency-regularization method for generalized Rayleigh quotient directivity specifications and two novel beamformer designs that optimize for maximum efficiency constant directivity (MECD) and maximum sensitivity constant directivity (MSCD). We derive fast converging and analytic solutions from their quadratic equality constrained quadratic program formulations. Experiments optimize generalized directivity index constrained beamformer designs for a full-band heterogeneous array.
\end{abstract}

\setlength{\belowdisplayskip}{6pt}
\setlength{\belowdisplayshortskip}{6pt}
\setlength{\abovedisplayskip}{6pt}
\setlength{\abovedisplayshortskip}{6pt}

\setlength{\skip\footins}{1mm}

\addtolength{\textfloatsep}{-0.2in}

%\setlength{\textfloatsep}{\baselineskip 0.2 \baselineskip -0.2\baselineskip}
%\setlength{\textfloatsep}{5pt}
%\setlength{\belowcaptionskip}{-10pt}

%\vspace*{-\baselineskip}
%\vspace*{-1.0mm}

\section{Introduction}
\label{SEC:INTRO}

Beamformer designs for loudspeaker arrays require constraints atypical to their microphone array counterpart due to greater variations in operating frequency characteristics across loudspeaker transducer elements, acoustic and electric power output limits, and placement restrictions. Traditional frequency-invariant beamforming designs for linear \cite{WARD_2001}, circular \cite{HUANG_2018}, spherical \cite{ZOTTER_2008}, differential \cite{WANG_2021, MIOTELLO_2023} homogeneous arrays often optimize for directivity factor and white-noise gain (WNG) \cite{BERKUM_2015}, \cite{CROCCO_2011}, and beam pattern fit \cite{PARRA_2006}. Such constraints are less applicable for heterogeneous loudspeakers and non-uniform arrays that jointly maximize acoustic/electrical power ratio in a listening window, and satisfy a generalized directivity index (GDI) \cite{LUO_2021_SPH}. The former extends WNG to  evaluation regions in spherical coordinates, and transducer dependent penalty factors per electrical power unit over frequency. The latter relaxes beam pattern targets to an acoustic power ratio over two evaluation regions defined by density functions. This work addresses both issues and is organized as follows:

Section \ref{SEC:SDO} introduces a novel regularization technique for the generalized Rayleigh quotient (GRQ) \cite{HORN_1990_MATRIX}  that penalizing both acoustic and electrical power for heterogeneous arrays without introducing explicit constraints in their formulations. Sections \ref{SEC:SDO:MECD}, \ref{SEC:SDO:MSCD} propose two maximal acoustic efficiency and sensitivity beamformer designs subject to constant GDI equality constraints, and present a pair of iterative and analytic solutions respectively via quadratic programming. Section \ref{SEC:SDO:QSE} solves for the secular equation sub-problem common to both beamformer formulations. Section \ref{SEC:EXP} shows several beamformer designs for a sample heterogeneous array and compares convergence rates of the iterative solutions. Section \ref{SEC:CONCLUSION} summarizes the theoretical results and experiments.

\section{Loudspeaker Directivity Optimization}
\label{SEC:SDO}

\begin{figure}[h]
  \centering
  {\includegraphics[height=2.6cm]{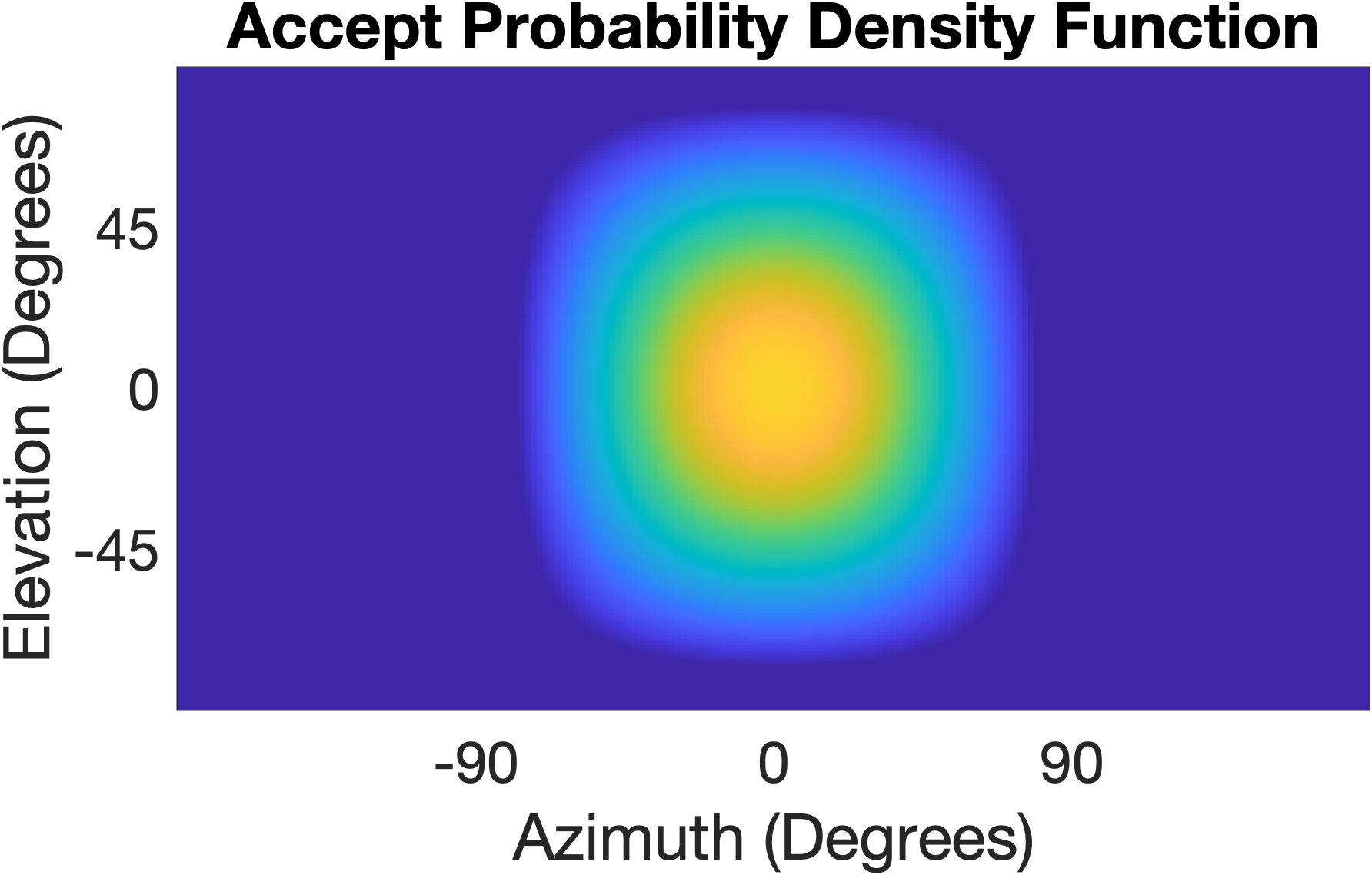}}
  \hspace{0.02cm}
  {\includegraphics[height=2.6cm]{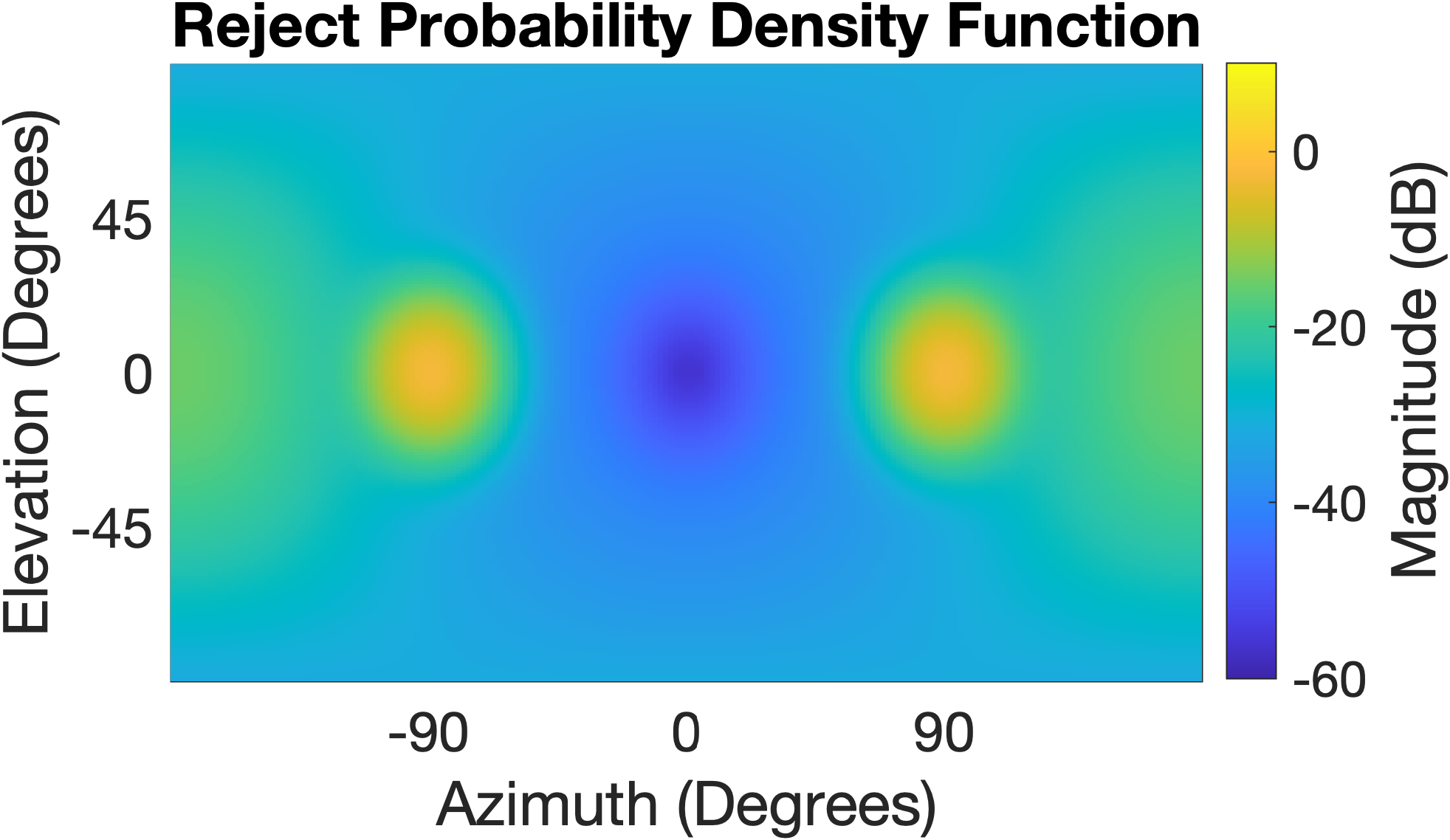}}
  
\vspace{-2.5mm}

\caption{Sample accept $f_A(\VEC{r})$ and reject $f_R(\VEC{r})$ density functions define forward-facing listening and side-facing reflection windows respectively.}

\vspace{-1.5mm}

\label{FIG:GRQ:PDF}
\end{figure}

The GRQ $\frac{\VEC{w}^H \MAT{A} \VEC{w}}{\VEC{w}^H \MAT{R} \VEC{w}}$ is the power ratio of a beamformer's responses over two domains specified by weights $\VEC{w}$ and so-called "accept" and "reject" covariance matrices $\MAT{A}$, and $\MAT{R}$ respectively. A beamformer's GDI equates to the GRQ for covariance matrices specified in the acoustic domain given by
 \begin{equation}
\begin{split}
\MAT{A} =  \EXPECTATION{\VEC{r} \small \sim f_A}{\VEC{d}(\VEC{r}) \VEC{d}^H(\VEC{r})  }, \,\,\, 
\MAT{R} =  \EXPECTATION{\VEC{r} \small \sim f_R}{\VEC{d}(\VEC{r}) \VEC{d}^H(\VEC{r})  },
\end{split}
\label{EQ:GRQ:COV}
\end{equation}
where the expectation samples steering vectors or anechoic frequency responses $\VEC{d}(\VEC{r})$  (conjugate transpose $\VEC{d}^H(\VEC{r})$) over the Cartesian unit-directions $\VEC{r}$. The probability density functions $f_A(\VEC{r})$, $f_R(\VEC{r})$ weight the acoustic power-responses over specifiable directions to boost and attenuate respectively as shown in Fig. \ref{FIG:GRQ:PDF}, and generalize the directivity index \cite{LUO_2021_SPH}. 
%%%%%%%
%%%%%
Maximizing GRQ therefore maximizes GDI by reducing GRQ into the normal Rayleigh quotient under a change of variables:
\begin{equation}
\begin{split}
G(\MAT{A}, \MAT{R}, \VEC{w}) & =  \frac{\VEC{w}^H \MAT{A} \VEC{w}}{ \VEC{w}^H \MAT{R} \VEC{w}} =
%%%%%%%
\frac{\VEC{x}^H \VEC{Q} \VEC{x}}{\VEC{x}^H \VEC{x}}, \quad 
%%%%%%%%%%
\VEC{x} = \MAT{L}^H \VEC{w}, \\
%%%%%%%%%
\MAT{Q} & = \MAT{L}^{-1} \MAT{A} \MAT{L}^{-H}, \quad 
%%%%%%%
\MAT{R}  = \MAT{L} \MAT{L}^H,
\end{split}
\label{EQ:GRQ:GRQ}
\end{equation}
where matrix $\MAT{L}$ is the lower-triangular Cholesky factor of $\MAT{R}$. The maximizer $\VEC{w}_* = \argmax_{\VEC{w}} G(\MAT{A}, \MAT{R}, \VEC{w}) = \MAT{L}^{-H} \VEC{v}$ of \eqref{EQ:GRQ:GRQ} constrained to the unit-circle $\VEC{w}^H \MAT{R} \VEC{w} = 1$ is therefore the eigenvector $\VEC{v}$ of the largest eigenvalue of matrix $\MAT{Q}$.

\begin{figure}[h]
  \centering
  {\includegraphics[width=0.84\linewidth]{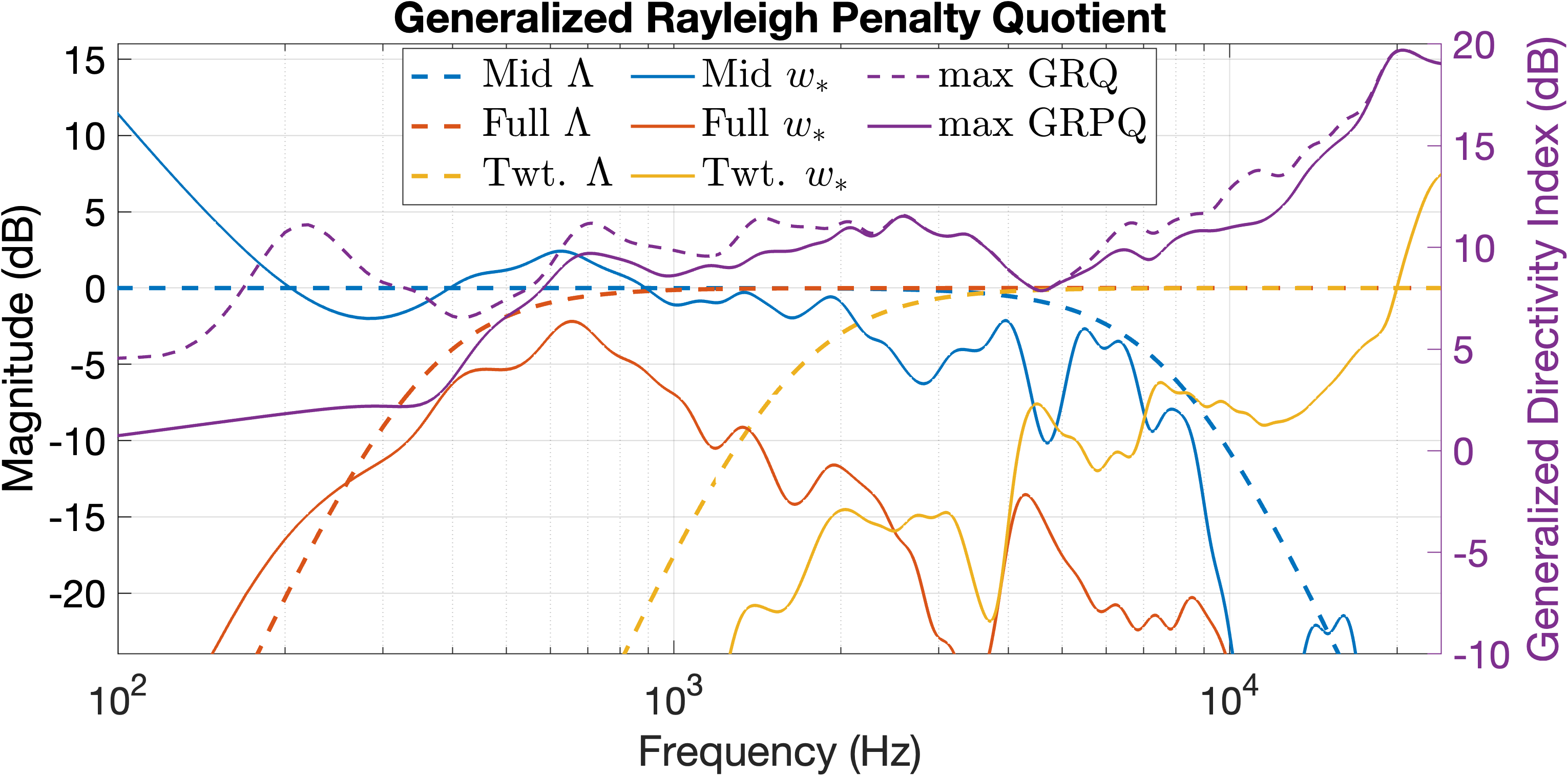}}
\vspace{-2.5mm}
\caption{Max GRQ upper-bounds max GRPQ for a real sample mid-range, full-range and tweeter array. The maximizer $\VEC{w}_*$ attenuates for frequencies outside each transducer's operating range curve specified by $\MAT{\Lambda}$.}
\vspace{-1.5mm}
\label{FIG:GRQ:XOVER}
\end{figure}

In heterogeneous transducer beamformer design, it is beneficial to mix both acoustic and electric rejection covariance matrices as to penalize components of $\VEC{w}$ outside each transducer's operating frequency ranges. Consider the following generalized Rayleigh penalty quotient (GRPQ) $G(\MAT{A}, \MAT{R} + \MAT{\Gamma} \MAT{\Sigma}, \VEC{w})$ which adds a diagonal positive-definite penalty matrix $\MAT{\Gamma} \MAT{\Sigma}$ to the denominator covariance matrix $\MAT{R}$ given by
\begin{equation}
\begin{split}
\MAT{\Gamma} = \DIAG{\DIAG{\MAT{R}}}, \quad
\MAT{\Sigma} = \DIAG{\sigma_1, \, \hdots, \, \sigma_N}, \quad 
\end{split}
\label{EQ:GRQ:GRPQ_GAMMA}
\end{equation}
where $0 \leq \sigma_n \leq \infty$ is unbounded. Inverting the sum of $\MAT{\Sigma}$ and the identity matrix $\MAT{I}$ gives the bounded weighting matrix
\begin{equation}
\begin{split}
\MAT{\Lambda}  & = \PR{\MAT{\Sigma} + \MAT{I}}^{-\frac{1}{2}} = \DIAG{\lambda_1, \, \hdots, \, \lambda_N}, \quad 0 \leq \lambda_n \leq 1, \\
\end{split}
\label{EQ:GRQ:GRPQ_W}
\end{equation}
which penalizes $\ABS{w_n}$ for smaller $\lambda_n$. Lowering $\lambda_n \rightarrow 0$ for frequencies outside transducer $n$'s operating range has the desired effect on the maximizer $\ABS{w_{n*}} \rightarrow 0$ and thus generalizes crossover designs to joint frequency-transducer weighted specifications of $\MAT{\Lambda}$ as shown in Fig. \ref{FIG:GRQ:XOVER}.
%%%%%%%
%%%%%%
For computation, $\MAT{\Lambda}$ also improves the condition number of $\MAT{\Gamma} \MAT{\Sigma}$ for maximizing \eqref{EQ:GRQ:GRQ} when combined with the following change of variables:
\begin{equation}
\begin{split}
\frac{\VEC{w}^H \MAT{A} \VEC{w}}{ \VEC{w}^H \PR{\MAT{R} + \MAT{\Gamma} \MAT{\Sigma} } \VEC{w}}  & =
%%%%%%%%%%%
\frac{\VEC{y}^H \MAT{\Lambda} \MAT{A} \MAT{\Lambda} \VEC{y}}{ \VEC{y}^H \MAT{\Lambda} \PR{\MAT{R} + \MAT{\Gamma} (\MAT{\Lambda}^{-2} - \MAT{I})} \MAT{\Lambda}  \VEC{y}}, \\
%%%%%%%%%
\end{split}
\label{EQ:GRQ:GRPQ}
\end{equation}
where $\VEC{y} = \MAT{\Lambda}^{-1} \VEC{w}$. The modified GRPQ in  \eqref{EQ:GRQ:GRPQ} is now well-conditioned and bounded above by GRQ as expressed by 
\begin{equation}
\begin{split}
G(\MAT{\Lambda} \MAT{A} \MAT{\Lambda}, \MAT{\hat{R}}, \VEC{y})  = G(\MAT{A}, \MAT{R}, \VEC{w})\, G(\MAT{\Lambda} \MAT{R} \MAT{\Lambda}, \MAT{\hat{R}}, \VEC{y}), \\
%%%%%%%
\MAT{\hat{R}}  = \MAT{\Lambda} \MAT{R} \MAT{\Lambda} + \MAT{\Gamma}(\MAT{I} - \MAT{\Lambda}^2) =  \MAT{\Gamma} + \MAT{\Lambda} \PR{\MAT{R} - \MAT{\Gamma} } \MAT{\Lambda}, \\
%%%%%%%%%%% 
\end{split}
\label{EQ:GRQ:GRPQ_R}
\end{equation}
where $\MAT{\hat{R}}$ attenuates non-diagonal entries of $\MAT{R}$ by $\MAT{\Lambda}^2$. 

\subsection{Maximum Efficiency Constant Directivity Beamformer}
\label{SEC:SDO:MECD}

In speaker array beamforming design, we can consider the following constant GRQ optimization $\argmax_{\VEC{w}} G(\MAT{C}, \MAT{I}, \VEC{w})$ s.t. $G(\MAT{A}, \MAT{R}, \VEC{w}) = \tau$, which maximizes an array's efficiency defined by the ratio of the acoustic output power for covariance matrix $\MAT{C}$, to electric output power s.t. constant GDI $\tau$. The constraint is feasible for $\tau$ between the smallest and largest generalized eigenvalues of $\MAT{A}$, $\MAT{R}$, and the solutions are not generally unique; consider diagonalizing matrix $\MAT{Q}$ in \eqref{EQ:GRQ:GRQ}:
\begin{equation}
\begin{split}
\tau = \frac{\VEC{x}^H \VEC{Q} \VEC{x}}{\VEC{x}^H \VEC{x}} = \frac{\VEC{z}^H \VEC{E} \VEC{z}}{\VEC{z}^H \VEC{z}}, \quad 
%%%%%%%
 \MAT{Q} = \MAT{V} \MAT{E} \MAT{V}^H, \,\,\,  \VEC{z} = \MAT{V}^H \VEC{x},
\end{split}
\label{EQ:GRQ:CRQ}
\end{equation}
where $\MAT{V} = \BK{\VEC{v}_1 \hdots \VEC{v}_N}$ is the column matrix of eigenvectors and $\VEC{e} = \DIAG{\MAT{E}} = \BK{e_1, \hdots, e_N}$ s.t. $e_n \leq e_{n+1}$ the ascending real-valued eigenvalues. For $\VEC{z}^H \VEC{z} = 1$ constrained to the unit-circle,  $\tau = \sum_{n=1}^N \ABS{z_n}^2 e_n$ is satisfied by multiple non-negative weighting of $\VEC{e}$ except at the extrema $\tau = \CR{e_1, e_N}$. For the open interval $e_1 < \tau < e_N$, we express GRQs \eqref{EQ:GRQ:GRQ}, \eqref{EQ:GRQ:CRQ} as the quadratic equality constraint $\VEC{w}^H \MAT{D} \VEC{w} = 0$ via substitution for the Hermitian matrix $\MAT{D}$, which is shown to be indefinite:
%%%%%%
\begin{equation}
\begin{split}
\MAT{D}  & = \MAT{A} - \tau \MAT{R} = \MAT{L} \MAT{V} \PR{\MAT{E} - \tau \MAT{I}} \MAT{V}^H \MAT{L}^H,
\end{split}
\label{EQ:GRQ:D_INDEF}
\end{equation}
where  $\VEC{w}_n  = \MAT{L}^{-H} \MAT{v}_n \, \Rightarrow \, \VEC{w}_n^H \MAT{D}\VEC{w}_n  = e_n - \tau$.

We now formally define the maximum efficiency constant directivity MECD$(\MAT{A}, \MAT{R}, \tau,  \MAT{C}, \VEC{w})$ beamformer design as a quadratic equality constrained quadratic program given by 
\begin{equation}
\begin{split}
%%%%%%%%%
\argmax_{\VEC{w}}  \VEC{w}^H  \MAT{C} \VEC{w} \quad \textrm{s.t. } \,\,  \VEC{w}^H \MAT{D} \VEC{w} = 0, \quad \VEC{w}^H \VEC{w} = 1,
%%%%%%%%%
\end{split}
\label{EQ:GRQ:CGRQ_GENERAL}
\end{equation}
where Hermitian matrix $\MAT{C}$ is the speaker covariances over a separate evaluation density $f_C(\VEC{r})$ as in \eqref{EQ:GRQ:COV}, and indefinite matrix $\MAT{D}$ constrains the GDI to $\tau$. The necessary conditions for optimality are expressed via the Lagrangian function $\mathcal{L}$:
%%%%%%%%%%%%%%%
\begin{equation}
\begin{split}
%%%%%%%%%
\mathcal{L}(\VEC{w}, \lambda, \mu) & =  \VEC{w}^H \MAT{C} \VEC{w} - \lambda (\VEC{w}^H  \VEC{D} \VEC{w}) - \mu (\VEC{w}^H \VEC{w} - 1), \\
%%%%%%%%%
\PDERIVFLAT{\mathcal{L}}{\VEC{w}}  & = \PR{\MAT{C}^T - \lambda \MAT{D}^T - \mu \MAT{I}} \VEC{w}^*, \\
%%%%%%%%
0 & = \PR{\PDERIVFLAT{\mathcal{L}}{\VEC{w}}}^* =  \PR{\MAT{C} - \lambda^* \MAT{D} - \mu^* \MAT{I}} \VEC{w}, \\
%%%%%%%%%
% &  \Rightarrow \lambda^* \MAT{D}\VEC{w}      =  \PR{\MAT{C} - \mu^* \MAT{I} } \VEC{w}, \\
%%%%%%%%%%
% \PDERIVFLAT{\mathcal{L}}{\VEC{w}^*}  & = \PR{\MAT{C} - \lambda \MAT{D} - \mu \MAT{I}} \VEC{w}, \quad \MAT{D} = \MAT{D}^H, \quad \MAT{C} = \MAT{C}^H, \\
%%%%%%%%%%
%\Rightarrow 0 & =  \PDERIVFLAT{\mathcal{L}}{\VEC{w}} + \PDERIVFLAT{\mathcal{L}}{\VEC{w}^*}    = 2 \PR{\MAT{C} - \Re\BK{\lambda} \MAT{D} - \Re\BK{\mu} \MAT{I} } \VEC{w}, \\
%%%%%%%%%%
 0 & = \PDERIVFLAT{\mathcal{L}}{\VEC{w}^*}    = \PR{\MAT{C} - \lambda \MAT{D} - \mu \MAT{I} } \VEC{w}, \\
%%%%%%%%%%
&  \Rightarrow
\Re\BK{\lambda} \MAT{D}\VEC{w}      =  \PR{\MAT{C} - \Re\BK{\mu} \MAT{I} } \VEC{w}, \\
\end{split}
\label{EQ:GRQ:CGRQ_HARD_LAGRANGE}
\end{equation}
where $\mu$ and $\lambda$ are the Lagrangian multipliers and it suffices to restrict both $\mu, \lambda  \in \mathcal{R}$ for Hermitian $\MAT{C}, \MAT{A}, \MAT{R}$. Substituting $\MAT{D} \VEC{w}$ in   \eqref{EQ:GRQ:CGRQ_HARD_LAGRANGE} into the constraints yields the critical points of $\mu$:
\begin{equation}
\begin{split}
%%%%%%
0 =  \PDERIVFLAT{\mathcal{L}}{\lambda} & =  \VEC{w}^H  \VEC{D} \VEC{w}  = \VEC{w}^H \PR{\MAT{C} - \mu \MAT{I} } \VEC{w},  \\
%%%%%%%%%%
\Rightarrow \mu & = \frac{\VEC{w}^H \MAT{C}\VEC{w}}{\VEC{w}^H \VEC{w}} = G(\MAT{C}, \MAT{I}, \VEC{w}), \\
%%%%%%%%%
\end{split}
\label{EQ:GRQ:CGRQ_HARD_LAGRANGE_SUB}
\end{equation}
which are bounded between smallest and largest eigenvalues of $\MAT{C}$. The stationary points of $\mathcal{L}(\VEC{w}, \lambda, \mu)$ can be found via iterative methods such the differential multipliers (DM) \cite{PLATT_CONS}, which at each iteration $k$ sets $\mu_k = G(\MAT{C}, \MAT{I}, \VEC{w}_k)$, performs a round of gradient ascent and descent on $\VEC{w}_k$, $\lambda_k$ via  \eqref{EQ:GRQ:CGRQ_HARD_LAGRANGE}, \eqref{EQ:GRQ:CGRQ_HARD_LAGRANGE_SUB} with fixed step-sizes $\alpha_{\VEC{w}}$, $\alpha_{\lambda}$ respectively, and lastly normalizes $\VEC{w}_k$ to the unit-circle.

% shown in Algorithm \eqref{ALG:GRQ:CGRQ_HARD}. This method combines renormalization of $\VEC{w}$ and the stationary point of $\mu$ at each iteration with the basic differential multiplier method from \cite{PLATT_CONS} whereby updating $\VEC{w}$ and $\lambda$ via gradient ascent and descent respectively converges to a local maximum.
%\begin{algorithm}
%	\KwRequire{Hermitian $\MAT{C}$, $\MAT{D} = \MAT{A} - \tau \MAT{R}$, $\min G(\MAT{A}, \MAT{R}, \VEC{w}) \leq \tau \leq \max G(\MAT{A}, \MAT{R}, \VEC{w}) $, $\alpha > 0$ step-size, $K$ iterations}
%    \KwResult{Solution and maximum to  \eqref{EQ:GRQ:CGRQ_GENERAL}}
%	$\VEC{w}_0, \lambda_0$ are initial points \\
%    \For{$k = 0$ to $K$}{
%        %execute phases\Comment{Mutation and crossover}
%		%%%%%%%%%%
%        $\mu = \frac{\VEC{w}^H_k \MAT{C} \VEC{w}_k}{\VEC{w}^H_k \VEC{w}_k}$, \quad Rayleigh Quotient\\
%        $\VEC{w}_* = \frac{\VEC{w}_k }{\NORM{\VEC{w}_k}}$, \quad Normalization \\
%        %%%%%%%%%%
%        $\VEC{w}_{k+1} = \VEC{w}_*  + \alpha \PR{\MAT{C} - \lambda_k \MAT{D} - \mu \MAT{I}} \VEC{w}_*$, \quad via  \eqref{EQ:GRQ:CGRQ_HARD_LAGRANGE} \\ 
%        		$\lambda_{k+1} = \lambda_{k} + \alpha \PR{\VEC{w}_*^H \MAT{D} \VEC{w}_* }$, \quad via  \eqref{EQ:GRQ:CGRQ_HARD_LAGRANGE_SUB} \\     
%		%%%%%%%%%%
%    }
%    \KwRet{$\VEC{w}_*$, $\mu$}
% \caption{MECD Lagrange Differential Multipliers}
% \label{ALG:GRQ:CGRQ_HARD}
%\end{algorithm}

We can greatly improve the convergence rate of the gradient method by directly solving for $\lambda_k$ via projecting $\VEC{w}_k$ along an unknown direction $\VEC{v}_*$ onto the nearest feasible point satisfying the quadratic equality constraints given by
\begin{equation}
\begin{split}
 \argmin_{\VEC{v}}  \VEC{v}^H \VEC{v} \quad \textrm{s.t. } \,\, \PR{\VEC{w}_k+\VEC{v}}^H \MAT{D} \PR{\VEC{w}_k + \VEC{v}} = 0,
\end{split}
\label{EQ:GRQ:CGRQ_HARD_PROJ}
\end{equation}
before normalizing the projection $\VEC{w}_k + \VEC{v}_*$ to the unit-circle.
%%%%%%%%%%%
\vspace{-2.5mm}
\begin{algorithm}
	\KwRequire{Hermitian $\MAT{C}$, $\MAT{D} = \MAT{A} - \tau \MAT{R}$, $\min G(\MAT{A}, \MAT{R}, \VEC{w}) \leq \tau \leq \max G(\MAT{A}, \MAT{R}, \VEC{w}) $, $\alpha > 0$ step-size, $K$ iterations, initial $\VEC{w}_0$}
    \KwResult{Solution and maximum to  \eqref{EQ:GRQ:CGRQ_GENERAL}}
    \For{$k = 1$ to $K$}{
		%%%%%%%%%%
		$\VEC{w}_* = \VEC{w}_{k-1} + \alpha \MAT{C} \VEC{w}_{k-1}$, \quad Gradient ascent \\
		$\VEC{v}_* = \textrm{Proj}(\VEC{w}_*, \MAT{D}) $, \quad Projection  \eqref{EQ:GRQ:CGRQ_HARD_PROJ} \\
	    $\VEC{w}_{k} = \frac{\VEC{w}_* + \VEC{v}_* }{\NORM{\VEC{w}_* + \VEC{v}_* }}$, \quad Normalization \\  	
		%%%%%%%%%%
    }
    \KwRet{$\VEC{w}_{K}$}
 \caption{MECD Projected Ascent}
 \label{ALG:GRQ:CGRQ_HARD_PROJ}
\end{algorithm}
\vspace{-2.5mm}
%%%%%%%%%%
This is expressed in the proposed MECD projected ascent (PA) Algorithm \eqref{ALG:GRQ:CGRQ_HARD_PROJ}, which iterates between moving $\VEC{w}$ along the gradient $\MAT{C}\VEC{w}$ of the objective in  \eqref{EQ:GRQ:CGRQ_GENERAL}, solving for the minimum norm projection vector $\VEC{v}_*$ for \eqref{EQ:GRQ:CGRQ_HARD_PROJ}, and normalizing the projection to $\PR{\VEC{w}_k + \VEC{v}_*}^H \PR{\VEC{w}_k + \VEC{v}_*}=1$.
%%%%%%%%
The minimum norm projection to  \eqref{EQ:GRQ:CGRQ_HARD_PROJ} is a variant of the quadratic constrained least-squares  problem in \cite{GANDER_1980} that we generalize for indefinite $\MAT{D}$. Its feasible surface is the intersection of two hyper-ellipsoids characterized by $\MAT{A}$ and $\tau \MAT{R}$ with necessary conditions satisfying the Lagrangian $\mathcal{L}$ and multiplier $\lambda$:
%%%%%%%%%%%%%%
\begin{equation}
\begin{split}
\mathcal{L}(\VEC{v}, \lambda) & =  \VEC{v}^H \VEC{v} - \lambda \PR{\VEC{w}_k + \VEC{v}}^H  \VEC{D} \PR{\VEC{w}_k + \VEC{v}}, \\
%%%%%%%
\PDERIVFLAT{\mathcal{L}}{\VEC{v}}  & =  \VEC{v}^* - \lambda  \MAT{D}^T \PR{\VEC{w}_k^* + \VEC{v}^*}, \\
%%%%%%%
0 &  = \PR{\PDERIVFLAT{\mathcal{L}}{\VEC{v}}}^*  =  \VEC{v} - \lambda^*  \MAT{D} \PR{\VEC{w}_k +  \VEC{v}}, \\
%%%%%%%%%
0 & = \PDERIVFLAT{\mathcal{L}}{\VEC{v}^*} = \VEC{v} - \lambda  \MAT{D} \PR{\VEC{w}_k + \VEC{v}}, \\
%%%%%%%%
\Rightarrow \VEC{v} &  = \Re \BK{\lambda}  \PR{ \MAT{I} - \Re \BK{\lambda} \MAT{D}}^{-1}  \MAT{D} \VEC{w}_k, 
\end{split}
\label{EQ:GRQ:CGRQ_HARD_PROJ_LAGRANGE}
\end{equation}
where it suffices to restrict $\lambda \in  \mathcal{R}$. It is useful to express  Hermitian $\MAT{D}  = \MAT{V} \MAT{E} \MAT{V}^H$  along the column matrix of eigenvectors $\MAT{V}$ and  diagonal matrix of real-valued eigenvalues $\MAT{E}$ so that $\lambda$ deflates the latter:
\begin{equation}
\begin{split}
%%%%%%%%
\VEC{v} & = \lambda  \MAT{V} \PR{ \MAT{I} - \lambda \MAT{E}}^{-1}   \MAT{E} \MAT{V}^H \VEC{w}_k, \\
%%%%%%%%%
\VEC{w}_k + \VEC{v} & = \PR{\VEC{I} + \lambda  \MAT{V} \PR{ \MAT{I} - \lambda \MAT{E}}^{-1}   \MAT{E} \MAT{V}^H } \VEC{w}_k\\
%%%%%%%%%
& = \VEC{V} \PR{\MAT{I} - \lambda \MAT{E}}^{-1} \VEC{V}^H \VEC{w}_k,
\end{split}
\label{EQ:GRQ:CGRQ_HARD_PROJ_LAGRANGE_V}
\end{equation}
where $\MAT{I} = \MAT{V} \MAT{V}^H = \MAT{V} \PR{\MAT{I} - \lambda \MAT{E}}^{-1}  \PR{\MAT{I} - \lambda \MAT{E}} \MAT{V}^H $.
Substituting $\VEC{v}$ into the constraints yields the critical points $\lambda$ which are the roots of a quadratic secular equation $S(\lambda)$ variant \cite{GOLUB_1973_EIG}:
\begin{equation}
\begin{split}
0 & =   S(\lambda) = \PDERIVFLAT{\mathcal{L}}{\lambda}  = \PR{\VEC{w}_k + \VEC{v}}^H \MAT{D} \PR{\VEC{w}_k + \VEC{v}}\\
%%%%%%%%
& = \VEC{u}^H \PR{\MAT{I} - \lambda \MAT{E}}^{-H} \MAT{E} \PR{\MAT{I} - \lambda \MAT{E}}^{-1} \VEC{u},
\end{split}
\label{EQ:GRQ:CGRQ_HARD_PROJ_LAGRANGE_SECULAR}
\end{equation}
where $\VEC{u} = \VEC{V}^H  \VEC{w}_k$ and the inner terms are diagonal matrices. $S(\lambda)$ is therefore a sum of rational quadratic functions.

We can further characterize the minimizer  $\PR{\VEC{v}_*, \lambda_*}$ via a slight modification of the first two theorems in the quadratic constrained least-squares solution \cite{GANDER_1980} to support indefinite $\MAT{D}$ for the squared norm (see Appendix derivations). Given a pair of critical points and solutions $\PR{\VEC{v}_1, \lambda_1}$ and $\PR{\VEC{v}_2, \lambda_2}$  satisfying  \eqref{EQ:GRQ:CGRQ_HARD_PROJ_LAGRANGE_V}, \eqref{EQ:GRQ:CGRQ_HARD_PROJ_LAGRANGE_SECULAR}, the difference of the squared norms between vectors $\VEC{v}_1, \VEC{v}_2$ relates to the difference in multipliers $\lambda_1 - \lambda_2$: 
%%%%%%%%%%%
\begin{equation}
\begin{split}
\frac{\lambda_1 - \lambda_2}{2} \PR{\VEC{v}_1 - \VEC{v}_2}^H \MAT{D} \PR{\VEC{v}_1 - \VEC{v}_2}  & = \VEC{v}_1^H \VEC{v}_1 -\VEC{v}_2^H \VEC{v}_2 ,
\end{split}
\label{EQ:GRQ:CGRQ_HARD_PROJ_LAGRANGE_SECULAR_THM1}
\end{equation}
and the squared norm of the vector difference $\VEC{v}_1 - \VEC{v}_2$ relates to the sum of multipliers $\lambda_1 + \lambda_2$:
\begin{equation}
\begin{split}
 \frac{\lambda_1 + \lambda_2}{2} \PR{\VEC{v}_1 - \VEC{v}_2}^H \MAT{D} \PR{\VEC{v}_1 - \VEC{v}_2}  &=  \PR{\VEC{v}_1 - \VEC{v}_2}^H \PR{\VEC{v}_1 - \VEC{v}_2} .
\end{split}
\label{EQ:GRQ:CGRQ_HARD_PROJ_LAGRANGE_SECULAR_THM2}
\end{equation}
%%%%%%
Unlike \cite{GANDER_1980}, the term $\PR{\VEC{v}_1 - \VEC{v}_2}^H \MAT{D} \PR{\VEC{v}_1 - \VEC{v}_2}$ may be negative for indefinite $\MAT{D}$ such that choosing the $\min \CR{\lambda_1, \lambda_2}$ via  \eqref{EQ:GRQ:CGRQ_HARD_PROJ_LAGRANGE_SECULAR_THM1} is insufficient for determining the smaller norm between $\VEC{v}_1, \VEC{v}_2$. Conversely,  \eqref{EQ:GRQ:CGRQ_HARD_PROJ_LAGRANGE_SECULAR_THM2} no longer ensures that there is at most one solution $\lambda < 0$. Taking the product of the left and right-hand sides of  \eqref{EQ:GRQ:CGRQ_HARD_PROJ_LAGRANGE_SECULAR_THM1}, \eqref{EQ:GRQ:CGRQ_HARD_PROJ_LAGRANGE_SECULAR_THM2} however gives
\begin{equation}
\begin{split}
 \frac{\lambda_1^2 - \lambda^2_2}{4}
%%%%%%%%%
& = \frac{\PR{\VEC{v}_1^H \VEC{v}_1 -\VEC{v}_2^H \VEC{v}_2} \PR{\VEC{v}_1 - \VEC{v}_2}^H \PR{\VEC{v}_1 - \VEC{v}_2}}{\PR{\PR{\VEC{v}_1 - \VEC{v}_2}^H \MAT{D} \PR{\VEC{v}_1 - \VEC{v}_2}}^2 }, 
\end{split}
\label{EQ:GRQ:CGRQ_HARD_PROJ_LAGRANGE_SECULAR_THM_PROD}
\end{equation}
which implies that picking the $\min \CR{\lambda_1^2, \lambda_2^2}$ gives the smaller norm between $\VEC{v}_1, \VEC{v}_2$. Therefore, the multiplier $\lambda_*$ nearest to $0$ and its corresponding vector $\VEC{v}_*$ after substitution into  \eqref{EQ:GRQ:CGRQ_HARD_PROJ_LAGRANGE_V} is the minimum norm solution under projection to  \eqref{EQ:GRQ:CGRQ_HARD_PROJ}. Methods for finding $
\lambda_*$ are discussed in Section \ref{SEC:SDO:QSE}.

We note that a semi-definite program (SDP) and its relaxation of MECD in  \eqref{EQ:GRQ:CGRQ_GENERAL} is also possible as the number constraints tightly bounds the rank of the solution \cite{SHAPIRO_RANK1, BARVINOK_RANK1, PATAKI_RANK1}. Formally, the MECD-SDP for unknown Hermitian and semi-definite matrix $\MAT{W} \succeq 0$ and the matrix trace operator $\TRACE{*}$ is given by 
\begin{equation}
\begin{split}
\argmax_{\VEC{W}}  \TRACE{\MAT{C} \MAT{W}} \quad  \textrm{s.t. } \,\,  \TRACE{\MAT{D} \MAT{W}} = 0, \quad \TRACE{\MAT{W}} = 1,
\end{split}
\label{EQ:GRQ:CGRQ_SDP}
\end{equation}
which for Hermitian matrices $\MAT{C}$, $\MAT{D}$ ensures a real-valued trace of two constraints, and a rank-$1$ solution $\MAT{W} = \VEC{w} \VEC{w}^H$. The interior-point method (IPM) therefore finds the maximizer within polynomial time as the problem is convex.

%We can seek the largest $\mu$ s.t. the generalized eigenvector $\VEC{w}$ and eigenvalue $\lambda$ pair for $\PR{C-\mu \MAT{I}} \VEC{w} = \lambda \MAT{D} \VEC{w}$ satisfies $\VEC{w}^H \MAT{D} \VEC{w} = 0$ and $\lambda \in \mathcal{R}$. However, the generalized eigenvalues $\lambda$ are not typically real as $\MAT{D}$ may not be positive definite.
%We therefore rearrange the terms where $\mu$ are the real eigenvalues of the Hermitian system:
%\begin{equation}
%\begin{split}
%\PR{\MAT{C} - \lambda \MAT{D}} \VEC{w} = \mu \VEC{w},
%\end{split}
%\label{EQ:GRQ:CGRQ_HARD_EIG}
%\end{equation}
%for unknown $\lambda$.

\subsection{Maximum Sensitivity Constant Directivity Beamformer}
\label{SEC:SDO:MSCD}

If the evaluation region is conventional\footnote{\vspace{-5.0mm}Speaker's acoustic response is measured at single point $\VEC{r}$ at $1$m distance}, then MECD maximizes the loudspeaker sensitivity \cite{DAVIS_2013_SOUND_SENSITIVITY} which we show has an analytic secular equation solution. Constraining the acoustic response to also be distortionless at the measurement point $\VEC{r}$ reduces MECD into the proposed maximum sensitivity constant directivity MSCD$(\MAT{A}, \MAT{R}, \tau,  \VEC{c}, \VEC{w})$ quadratic program with both quadratic and linear constraint beamformer:
%%%%%%%
\begin{equation}
\begin{split}
%%%%%%%%%
\argmin_{\VEC{w}} \VEC{w}^H \VEC{w} &  \quad \textrm{s.t. } \,\, \VEC{w}^H \MAT{D} \VEC{w}  = 0, \quad \MAT{c}^H \VEC{w} = 1, \\
%%%%%%%%%
%\MAT{D} & = \MAT{A} - \tau  \MAT{R}, \quad \VEC{c} = \MAT{d}(\VEC{r}),
%%%%%%%%%
\end{split}
\label{EQ:GRQ:CGRQ_SIMPLE}
\end{equation}
whereby  $\VEC{c} = \MAT{d}(\VEC{r})$  and the acoustic output at $\VEC{r}$ is constrained to  unity $\VEC{c}^H \VEC{w} = 1$ and therefore has unit acoustic power $\VEC{w}^H \MAT{C} \VEC{w} = 1$ for $\MAT{C} = \VEC{d}(\VEC{r}) \VEC{d}(\VEC{r})^H$. This problem is a variant of \cite{GOLUB_1973_EIG, GANDER_1989_EIG} where the latter's quadratic equality is set to unity instead of zero.
%%%%%%%%%%%%%
The MSCD extrema are found at the critical points of the Lagrangian function $\mathcal{L}$ and multipliers $\mu$ and $\lambda$: 
\begin{equation}
\begin{split}
%%%%%%%%%
\mathcal{L}(\VEC{w}, \lambda, \mu) & =  \VEC{w}^H \VEC{w} - \lambda (\VEC{w}^H  \VEC{D} \VEC{w}) \\ & - \mu (\VEC{c}^H \VEC{w} - 1) - \mu^* (\VEC{c}^T \VEC{w}^* - 1), \\
%%%%%%%%%
 \PDERIVFLAT{\mathcal{L}}{\VEC{w}}  & = \PR{\MAT{I} - \lambda \MAT{D}^T} \VEC{w}^* - \mu \VEC{c}^*, \\
 %%%%%%%%%
 0 & = \PR{%%%%%%%%%
 \PDERIVFLAT{\mathcal{L}}{\VEC{w}} }^*  = 
\PR{\MAT{I} - \lambda^* \MAT{D}} \VEC{w} - \mu^* \VEC{c} , \\
%%%%%%%%  
0 & =  \PDERIVFLAT{\mathcal{L}}{\VEC{w}^*} =  \PR{\MAT{I} - \lambda \MAT{D}} \VEC{w} - \mu^* \VEC{c}, \\
%%%%%%%%%
% \Rightarrow 0 & = \PDERIVFLAT{\mathcal{L}}{\VEC{w}}  + \PDERIVFLAT{\mathcal{L}}{\VEC{w}^*} = 2 \PR{\MAT{I} - \Re\BK{\lambda} \MAT{D} } \VEC{w} - \mu^* \VEC{c}, \\
%%%%%%%%%
&  \Rightarrow \VEC{w}   = \mu^* \PR{\MAT{I} - \Re\BK{\lambda} \MAT{D}}^{-1} \VEC{c},
\end{split}
\label{EQ:GRQ:CGRQ_SIMPLE_LAGRANGE}
\end{equation}
where it suffices to restrict $\lambda \in \mathcal{R}$ for Hermitian $\MAT{A}, \MAT{R}$. Substituting $\VEC{w}$ into the constraints yields the critical points:
\begin{equation}
\begin{split}
%%%%%%
0 = \PDERIVFLAT{\mathcal{L}}{\mu} & =  \VEC{c}^H \VEC{w} - 1  \,\,\, 
%%%%%%%%%%%%%%
\Rightarrow  \,\,\, \mu^* =  \PR{\VEC{c}^H \PR{\MAT{I} - \lambda \MAT{D}}^{-1} \VEC{c}}^{-1}, \\
%%%%%%%%%
\Rightarrow \quad &   \VEC{w}   = \frac{\PR{\MAT{I} - \lambda \MAT{D}}^{-1} \VEC{c}}{\VEC{c}^H \PR{\MAT{I} - \lambda \MAT{D}}^{-1} \VEC{c}},  \\
%%%%%%%%%
0 =  \PDERIVFLAT{\mathcal{L}}{\lambda} & =  \VEC{w}^H  \VEC{D} \VEC{w}   = \VEC{c}^H \PR{\MAT{I} - \lambda \MAT{D}}^{-H} \MAT{D}  \PR{\MAT{I} - \lambda \MAT{D}}^{-1} \VEC{c},
\end{split}
\label{EQ:GRQ:CGRQ_SIMPLE_LAGRANGE_SUB}
\raisetag{30pt}
\end{equation}
which restricts $\mu \in \mathcal{R}$  as the eigenvalues of $\MAT{I} - \lambda \MAT{D}$ are also real. The critical points $\lambda$ are the real-valued roots of  $S(\lambda) = \PDERIVFLAT{\mathcal{L}}{\lambda}$ in  \eqref{EQ:GRQ:CGRQ_SIMPLE_LAGRANGE_SUB}, analogous to that of  \eqref{EQ:GRQ:CGRQ_HARD_PROJ_LAGRANGE_SECULAR} and given by
\begin{equation}
\begin{split}
S(\lambda) & = \VEC{w}^H  \VEC{D} \VEC{w}  = \VEC{u}^H (\MAT{I} - \lambda \MAT{E})^{-H} \MAT{E} (\MAT{I} - \lambda \MAT{E})^{-1}   \VEC{u},   \\
%%%%%%%%%
\MAT{D}  & = \MAT{V} \MAT{E} \MAT{V}^H, \quad \MAT{I} - \lambda \MAT{D}  = \MAT{V}(\MAT{I} - \lambda \MAT{E}) \MAT{V}^H,    \\
%%%%%%%%
\end{split}
\label{EQ:GRQ:CGRQ_SECULAR}
\end{equation}
where $\VEC{u} =  \VEC{V}^H \VEC{c}$, column eigenvector matrix $\MAT{V}$ and  diagonal matrix of real-valued eigenvalues $\MAT{E}$ of $\MAT{D}$. The minimizer $\VEC{w}_*$ is therefore found at the root $\lambda_*$  nearest to $0$.

%Substituting the real-valued roots  $\lambda_*$ into $\VEC{w}$ in  \eqref{EQ:GRQ:CGRQ_SIMPLE_LAGRANGE_SUB} yields the extrema of $\VEC{w}^H \VEC{w}$.

%The GRPQ and MSCD results are combined for heterogeneous loudspeaker beamforming designs. Applying the weighting matrix $\MAT{\Lambda}$ in  \eqref{EQ:GRQ:GRPQ_W} to $\VEC{w}$ via the change of variables $\VEC{y} = \MAT{\Lambda}^{-1} \VEC{w}$ and preserving the original constraints in  \eqref{EQ:GRQ:CGRQ_SIMPLE} yields the penalized MSCD or  $\textrm{PMSCD}(\MAT{\Lambda}, \MAT{A}, \MAT{R}, \tau, \VEC{c}, \VEC{w}) = \textrm{MSCD}(\MAT{\Lambda}^H\MAT{A} \MAT{\Lambda}, \MAT{\Lambda}^H\MAT{R} \MAT{\Lambda}, \tau, \MAT{\Lambda}^H \VEC{c}, \VEC{y} )$ beamformer. (Note that $\tau$ range is more restricted)

\section{Quadratic Secular Equation Root-Finding}
\label{SEC:SDO:QSE}
%%%%%%%%%%%
The secular equations $S(\lambda)$ in  \eqref{EQ:GRQ:CGRQ_HARD_PROJ_LAGRANGE_SECULAR}, \eqref{EQ:GRQ:CGRQ_SECULAR} can be expressed in rational form w.r.t. reciprocal eigenvalue poles $b_n = e_n^{-1}$:
\begin{equation}
\begin{split}
%%%%%%%%
S(\lambda) & = \sum_{n=1}^N \frac{u_n u_n^* e_n}{\PR{1 - \lambda e_n}^2} = \sum_{n=1}^N \frac{a_n b_n  }{\PR{\lambda - b_n }^2}, 
\end{split}
\label{EQ:GRQ:CGRQ_SECULAR_ALGEBRA}
\end{equation}
where  $u_n$ and $e_n$ are the $n$th elements of $\VEC{u}$ and $\DIAG{\MAT{E}}$  from  \eqref{EQ:GRQ:CGRQ_HARD_PROJ_LAGRANGE_SECULAR}, \eqref{EQ:GRQ:CGRQ_SECULAR} respectively, and $a_n = \ABS{u_n}^2$. It is also useful to express $S(\lambda)  = S_{\minus}(\lambda) + S_{\plus}(\lambda) $ and its first derivative in terms of the set of negative and positive poles $B_{\minus} = \CR{b_n | \, b_n < 0}$ and $B_{\plus} = \CR{b_n | \, b_n > 0}$ respectively given by
\begin{equation}
\begin{split}
S_{\minus}(\lambda) & = \hspace{-0.15cm} \sum_{b_n \in B_{\minus}} \frac{a_n b_n  }{\PR{\lambda - b_n }^2}, \,\,\,\,
%%%%%%
\PDERIV{S_{\minus}(\lambda)}{\lambda}  = \hspace{-0.1cm} \sum_{b_n \in B_{\minus}} \frac{2 a_n \ABS{b_n}  }{\PR{\lambda - b_n}^3}, \\
%%%%%%%
S_{\plus}(\lambda) &  = \hspace{-0.15cm}  \sum_{b_n \in B_{\plus}} \frac{a_n b_n  }{\PR{\lambda - b_n }^2}, \,\,\,\,
%%%%%%%%
\PDERIV{S_{\plus}(\lambda)}{\lambda}  =  \hspace{-0.1cm} \sum_{b_n \in B_{\plus}} \frac{-2 a_n b_n  }{\PR{\lambda - b_n}^3}.
\end{split}
\label{EQ:GRQ:CGRQ_SECULAR_ALGEBRA_DERIVS}
\end{equation}
Unlike the linear and quadratic secular equations in \cite{GOLUB_1973_EIG}, $S(\lambda)$ is not monotonic in most intervals between consecutive poles due to the presence of different signed $b_n$ in the numerator terms. The one exception is the tightest interval bounding $0$ which contains the smallest magnitude negative and positive poles $b_{\minus} = \max \PR{B_{\minus}}$, $b_{\plus} = \min \PR{B_{\plus}}$ respectively. This is a consequence of  \eqref{EQ:GRQ:CGRQ_SECULAR_ALGEBRA_DERIVS} where the first derivatives given by
\begin{equation}
\begin{split}
\PDERIV{S_{\minus}(\lambda > b_{\minus})}{\lambda} > 0, \quad 
%%%%%%%%
\PDERIV{S_{\plus}(\lambda < b_{\plus})}{\lambda} > 0,
\end{split}
\label{EQ:GRQ:CGRQ_SECULAR_ALGEBRA_DERIVS_BOUNDS}
\end{equation}
in the intervals outside these poles are bounded and positive.
%%%%%%%
\begin{figure}[h]
\vspace{-2.5mm}

  \centering
  {\includegraphics[width=0.8\linewidth]{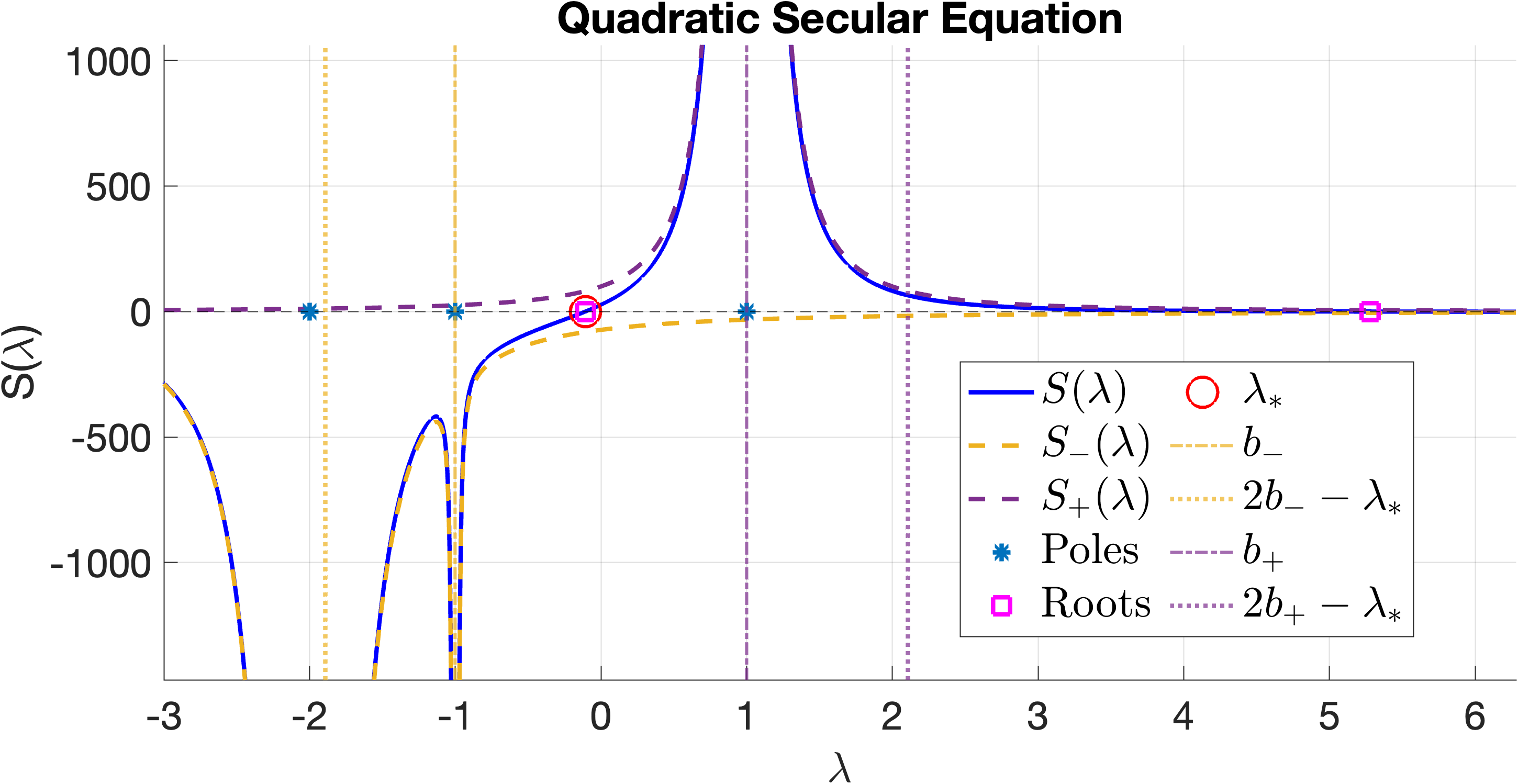}}

\vspace{-2.5mm}

\caption{$S(\lambda)$ has positive pole $B_{\pluscap} = \CR{1}$, negative poles $B_{\minuscap} = \CR{-2, -1}$, and two roots. The monotonic increasing $S(b_{\minuscap} < \lambda < b_{\pluscap})$  bounds the root $\lambda_*$ nearest to $0$ and restricts other roots to outside $\lambda_*$ reflected across $b_{\minuscap}$, $b_{\pluscap}$.}
\label{FIG:GRQ:SECULAR}
\end{figure}
%%%%%%%%%%
Moreover,  we show as in Fig. \ref{FIG:GRQ:SECULAR} that the intersecting interval $b_{\minus} < \lambda <  b_{\plus}$ contains exactly one real-valued root $\lambda_*$ that is also the root nearest to $0$ and therefore equivalent to the critical point and solutions to  \eqref{EQ:GRQ:CGRQ_HARD_PROJ}, \eqref{EQ:GRQ:CGRQ_SIMPLE}. The proof is as follows:

The function  $S(b_{\minus} < \lambda <  b_{\plus})$  where $S(\lambda = b_{\minus})=-\infty$, $S(\lambda = b_{\plus}) = +\infty$ is continuous and so the intermediate value theorem guarantees the existence of at least one real-valued root in the interval. Its first derivative $\PDERIV{S( b_{\minus} <  \lambda < b_{\plus})}{\lambda} > 0$ is strictly positive in the interval via \eqref{EQ:GRQ:CGRQ_SECULAR_ALGEBRA_DERIVS_BOUNDS}  and therefore contains exactly one root $\lambda_*$ that can easily be found via bracketing methods such as \cite{BRENT_1971}. To show that $\lambda_*$ is indeed the nearest to $0$, observe that reflecting $\lambda$ over $b_{\minus}$ or $b_{\plus}$ in the interval moves $\lambda$ closer to the remaining poles in $B_-$, $B_+$ respectively:
\begin{equation}
\begin{split}
\ABS{\lambda - b_n} & > \ABS{2 b_{\minus} - \lambda - b_n}, \quad b_n \in B_{\minus} \setminus \CR{b_{\minus}}, \,\,\, \lambda > b_{\minus}, \\
%%%%%%%%%%%%
\ABS{\lambda - b_n} & > \ABS{2 b_{\plus} - \lambda - b_n}, \quad b_n \in B_{\plus} \setminus \CR{b_{\plus}}, \,\,\,  \lambda < b_{\plus},
%%%%%%
\end{split}
\label{EQ:GRQ:CGRQ_SECULAR_ALGEBRA_BOUNDED_REFL}
\end{equation}
which with  \eqref{EQ:GRQ:CGRQ_SECULAR_ALGEBRA_DERIVS_BOUNDS} places bounds on  $S_{\minus}(\lambda)$, $S_{\plus}(\lambda)$ given by
\begin{equation}
\begin{split}
%%%%%%%%%%%%%%%%
S_{\minus}(\lambda)  > S_{\minus}\PR{2 b_{\minus} -  \lambda},  \,\,\, S_{\minus}(\lambda)  < S_{\minus}\PR{2 b_{\plus} -  \lambda},  &   \,\,\, \lambda > b_{\minus}, \\
%%%%%%%%%%%%%%%%
S_{\plus}(\lambda) < S_{\plus}\PR{2 b_{\plus} -  \lambda}, \,\,\, S_{\plus}(\lambda) > S_{\plus}\PR{2 b_{\minus} -  \lambda}, &  \,\,\,  \lambda < b_{\plus}, \\
%%%%%%%%%%%%%%%
\end{split}
\label{EQ:GRQ:CGRQ_SECULAR_ALGEBRA_BOUNDED_INEQ}
\end{equation}
and as a result induces bounds on $S(\lambda)$ after summation:
\begin{equation}
\begin{split}
S(2 b_{\minus} - \lambda)  < S(\lambda) < S(2 b_{\plus} - \lambda), & \quad  b_{\minus} < \lambda <  b_{\plus}, \\
%%%%%%%%%%
\Rightarrow \, S(2 b_{\minus} - \lambda) < S(\lambda) < 0, & \quad  b_{\minus}  < \lambda < \lambda_{*}, \\
%%%%%%%%%
\Rightarrow \, 0 < S(\lambda) < S(2 b_{\plus} - \lambda), & \quad  \lambda_{*} < \lambda <  b_{\plus}.
\end{split}
\label{EQ:GRQ:CGRQ_SECULAR_ALGEBRA_BOUNDED_INEQ_FIN}
\end{equation}
%%%%%%%%%%%
This implies there are no roots in the regions reflected across $b_{\minus}$ and $b_{\plus}$ given by $2b_{\minus} - \lambda_* < \lambda < b_{\minus}$ and   $b_{\plus} < \lambda < 2 b_{\plus} - \lambda_* $ respectively and that $\lambda_*$ must be nearest to $0$.

\section{Experiments}
\label{SEC:EXP}

\begin{figure}[h]
\vspace{-2.5mm}

  \centering
    {\includegraphics[width=0.85\linewidth]{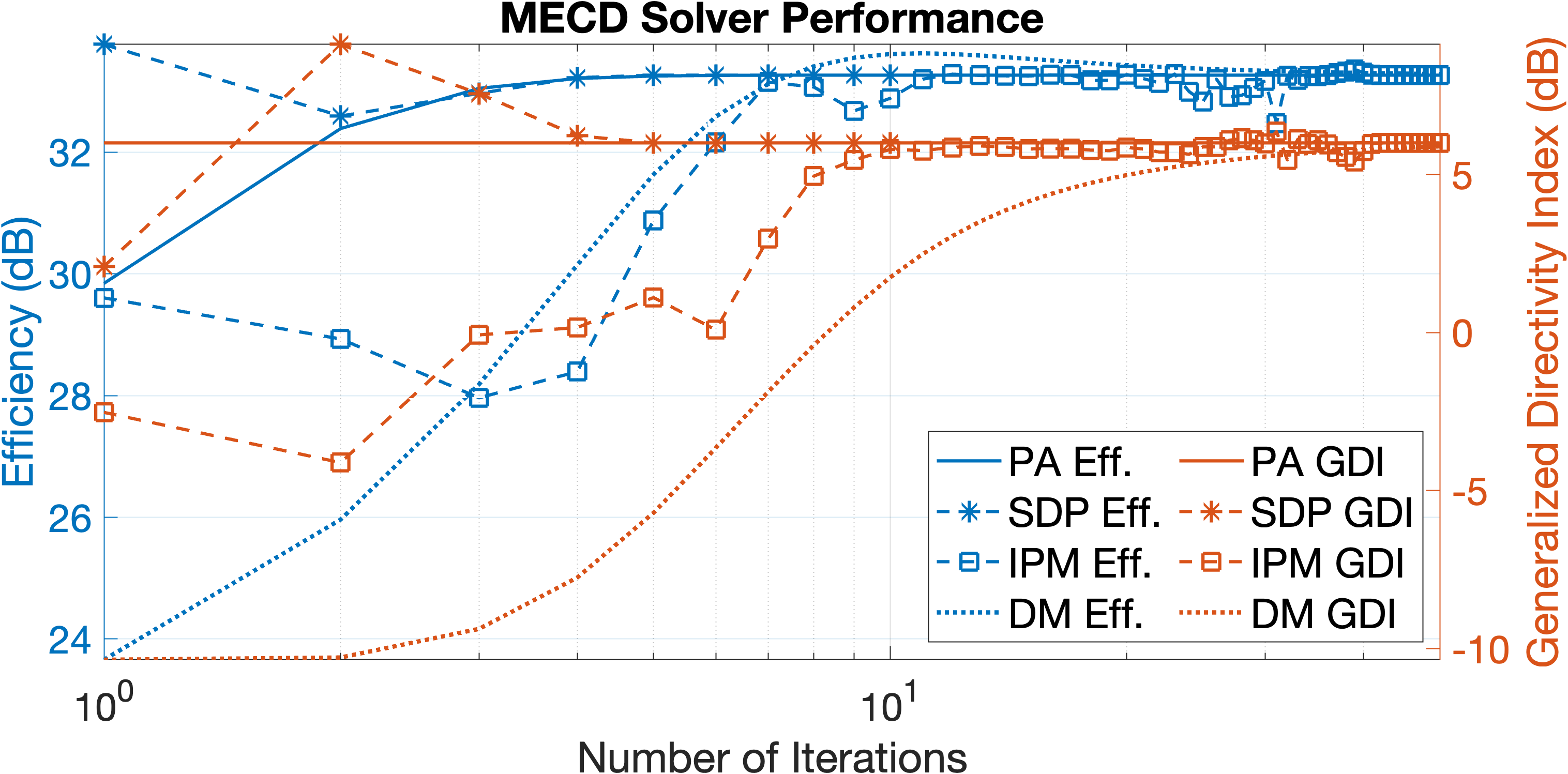}}
       
\vspace{-2.5mm}

\caption{MECD solver convergence rates for efficiency  $G(\MAT{C}, \MAT{I}, \VEC{w})$ objective and constant GDI $G(\MAT{A}, \MAT{R}, \VEC{w}) = 6 \textrm{ dB}$ constraint are ranked PA $(\alpha=1, \VEC{w}_0 = \VEC{1})$ $>$ SDP $>$ IPM $>$ DM ($\alpha_{\VEC{w}} =  1e-2$, $\alpha_{\lambda} = 1e-3$ step-sizes).}

\vspace{-1.5mm}

\label{FIG:EXP:SOLVER}
\end{figure}

We first compare MECD solver performance for $N=8$ elements with randomized complex-valued covariances $\MAT{A}, \MAT{R}, \MAT{C}$ in Fig. \ref{FIG:EXP:SOLVER}.  Our PA method reduces the number of iterations to convergence of the baseline DM method from $50$ to $5$. Larger step-sizes in the latter cause oscillations between the objective and constraint, and also exhibit to a lesser degree in Matlab's \verb+fmincon+ IPM solver. \verb+SDPT3+ \cite{SDPT3} for SDP shows similar convergence rates of the objective to PA but requires more compute in solving for $N^2$ number of variables instead of $N$.

\begin{figure}[h]

\vspace{-2.5mm}

  \centering
%  {\includegraphics[height=2.5cm]{figs/hdir_max_GRQ.png}}
%  {\includegraphics[height=2.5cm]{figs/hdir_max_GRPQ.png}}
%  
%  	\vspace{1.5mm}
%  	
%  {\includegraphics[height=2.5cm]{figs/hdir_MECD_GRQ.png}}
%  {\includegraphics[height=2.5cm]{figs/hdir_MECD_GRPQ.png}}
%
%	\vspace{1.5mm}
%  
%  {\includegraphics[height=3.2cm]{figs/hdir_MSCD_GRQ.png}}
%  {\includegraphics[height=3.2cm]{figs/hdir_MSCD_GRPQ.png}}
  
  {\includegraphics[height=2.35cm]{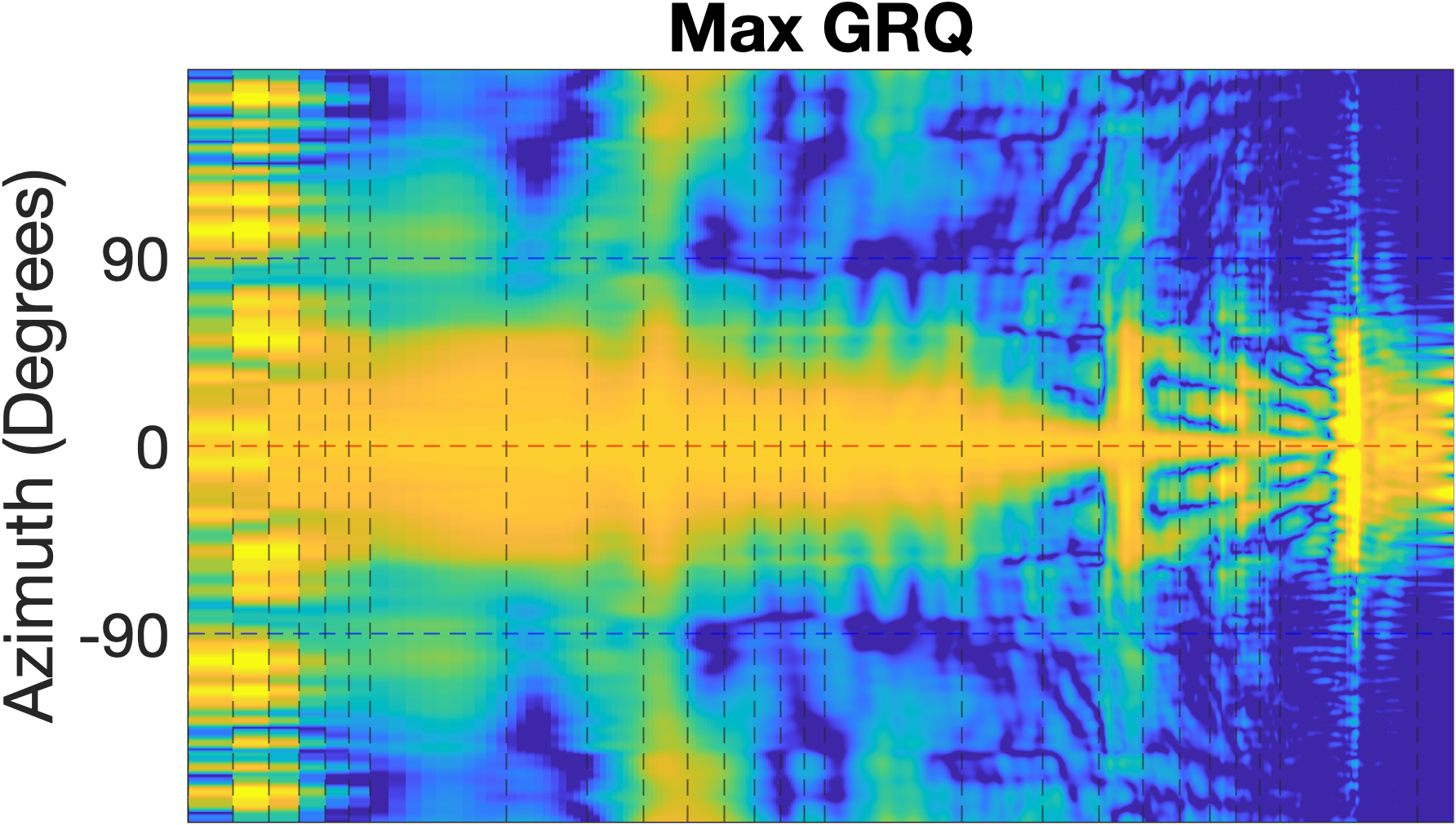}}
  {\includegraphics[height=2.35cm]{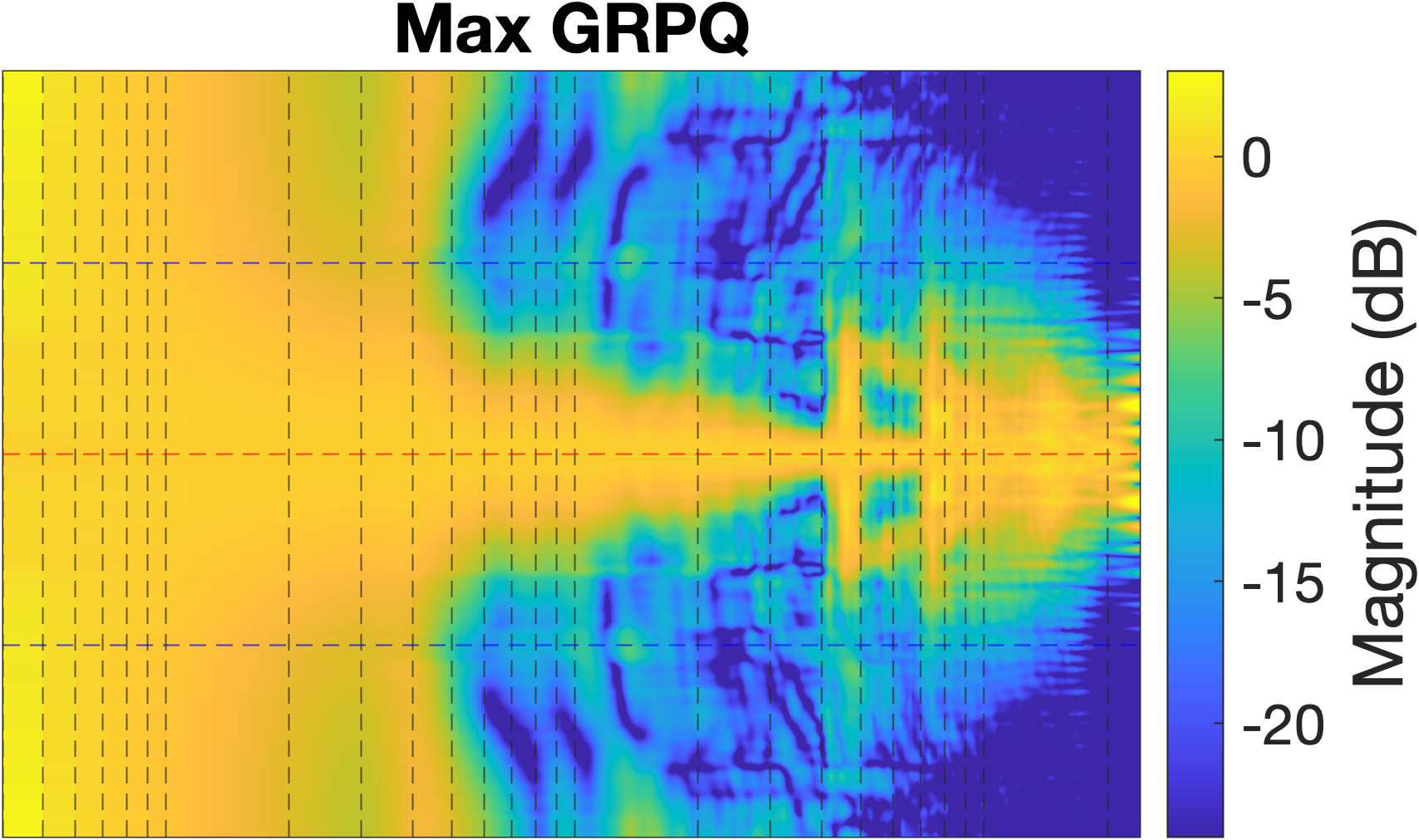}}
  
  	\vspace{1.5mm}
  	
  {\includegraphics[height=2.35cm]{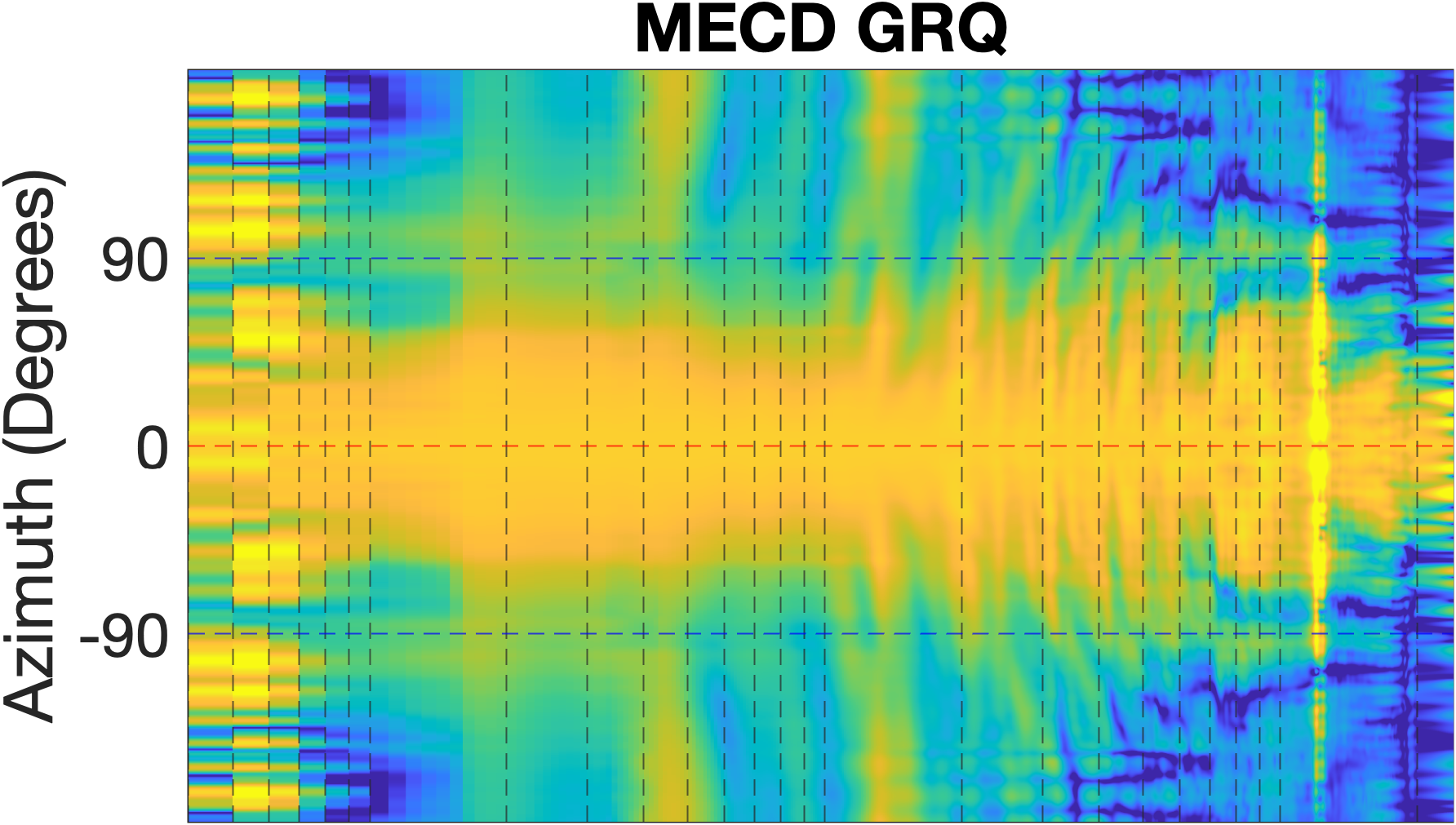}}
  {\includegraphics[height=2.35cm]{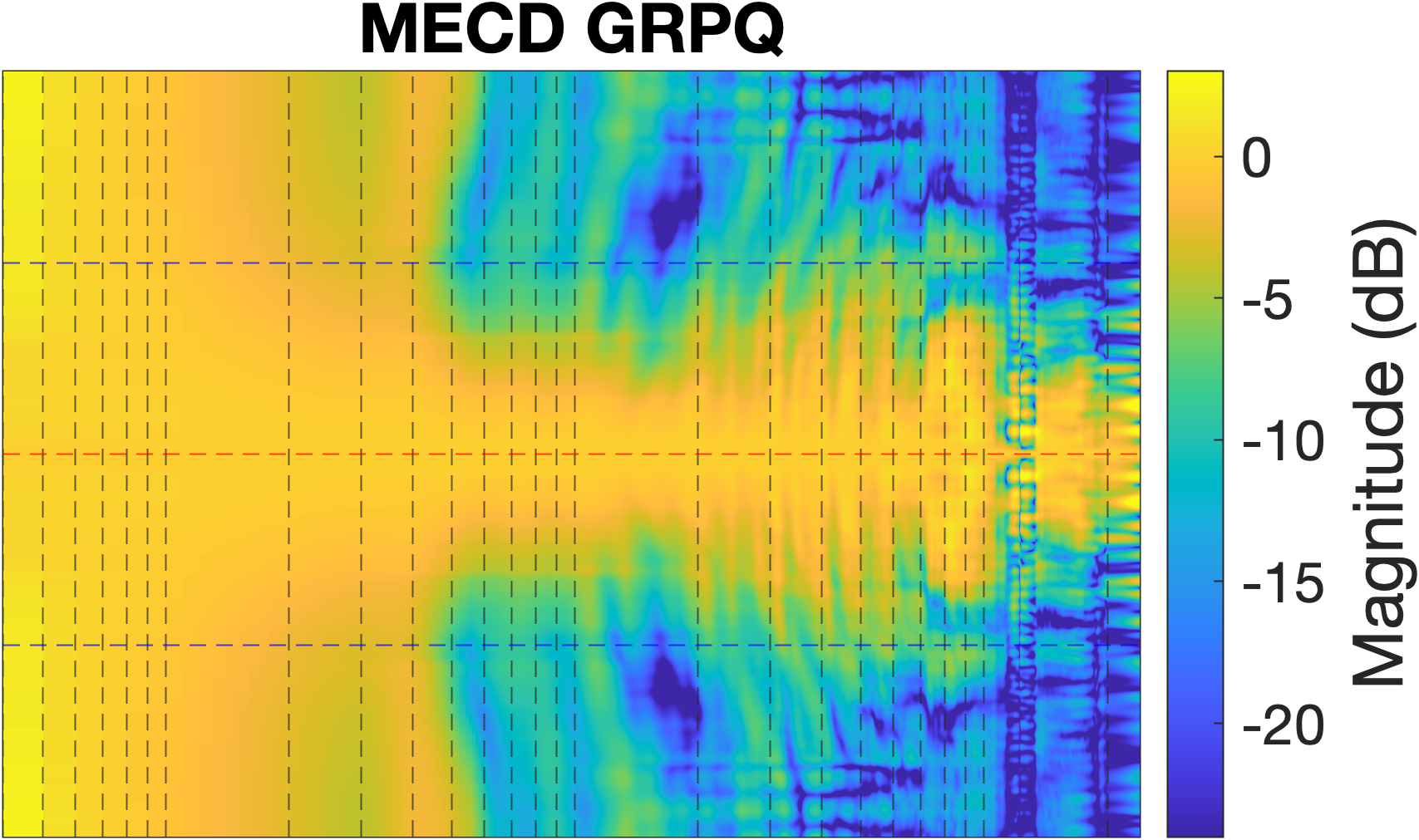}}

	\vspace{1.5mm}
  
  {\includegraphics[height=3.0cm]{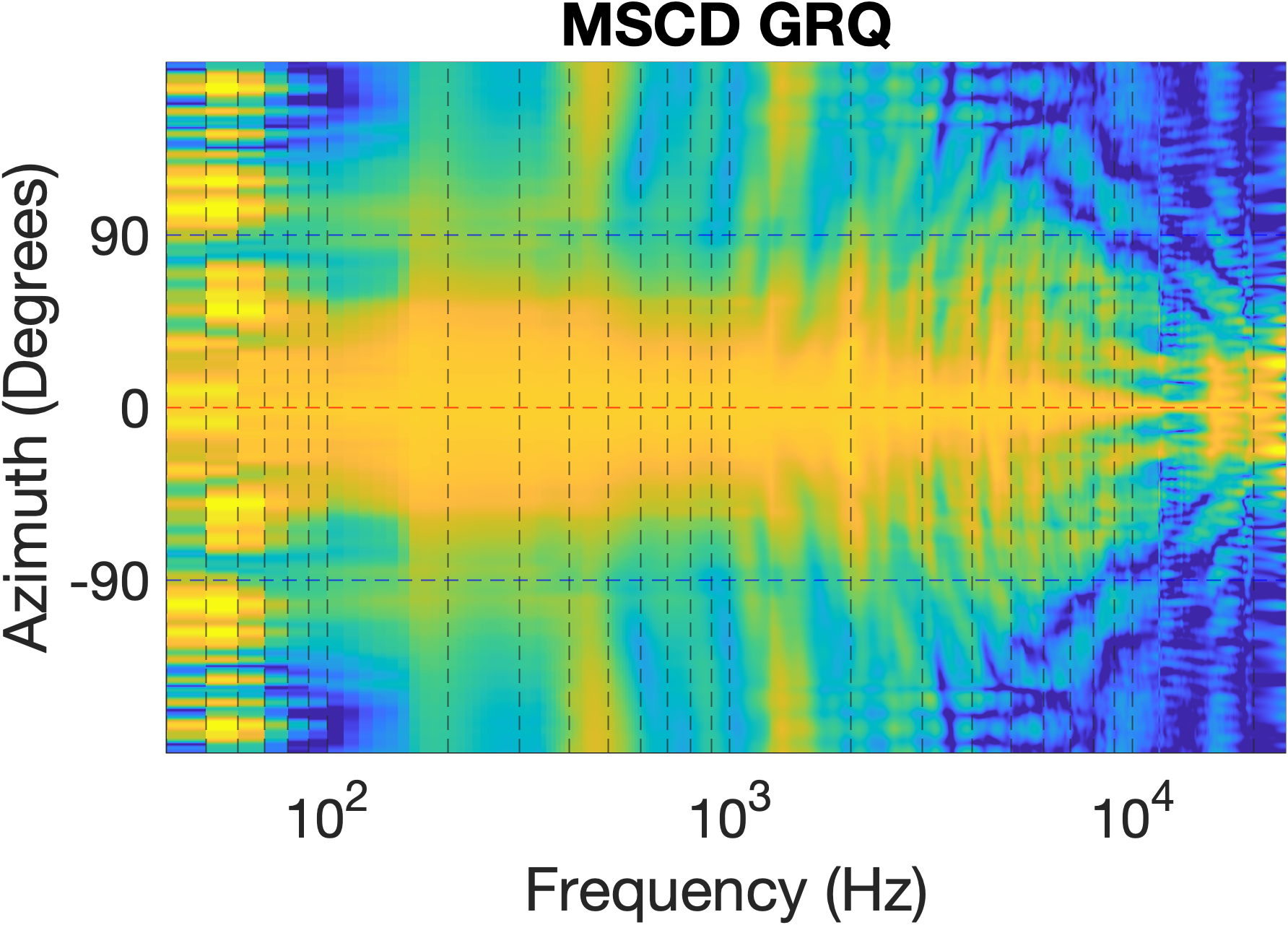}}
  {\includegraphics[height=3.0cm]{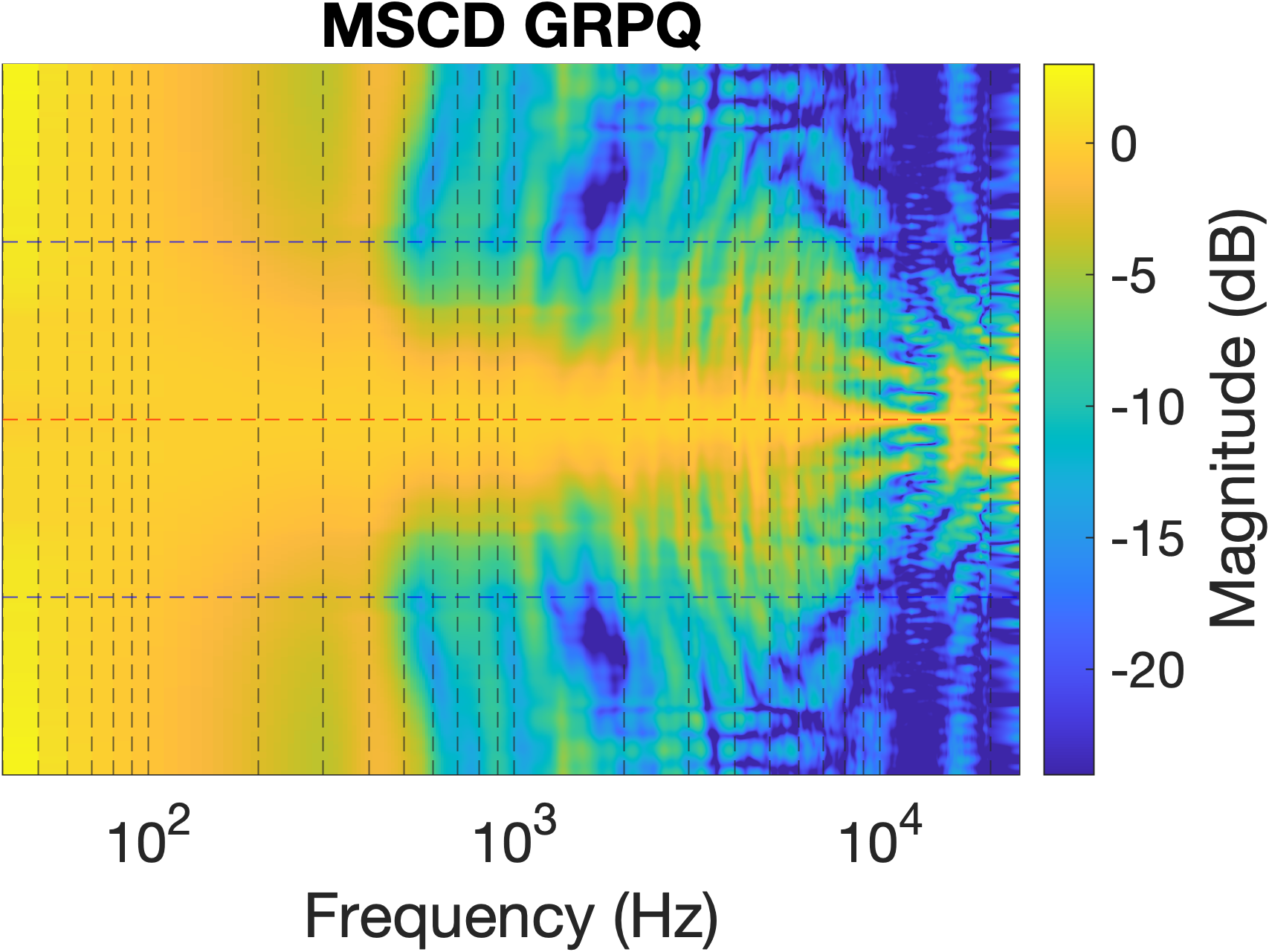}}
  
\vspace{-2.5mm}

\caption{Contour plots compare GRQ to GRPQ beam patterns constrained to each transducer's operating ranges across frequency on the horizontal plane. }

\vspace{-1.0mm}

\label{FIG:EXP:HDIR}
\end{figure}

We then compare GRQ to GRPQ beamformer designs for a measured sample $N=3$ mid-range, full-range, and tweeter array with $\MAT{A}, \MAT{R}, \MAT{C}=\MAT{A}$ integrated over densities from Fig. \ref{FIG:GRQ:PDF}, frequency weighting $\Lambda$ from Fig. \ref{FIG:GRQ:XOVER}, and constant GDI $\min \CR{6, \, 10 \log_{10} \max (\tau) }  \textrm{ dB} $ for MECD and MSCD. GRPQ beam patterns shown in Fig. \ref{FIG:EXP:HDIR} exhibit more regular responses inline with expected transducer operating frequency ranges; directivity grows increasingly omni-directional in low-frequency as only the mid-range is active. MECD beam patterns exhibit less lobing than MSCD and their maximum GDI counterparts. 

\section{Conclusion}
\label{SEC:CONCLUSION}
We introduced an electrical power penalty term to GRQ for regularizing frequency-dependent heterogeneous speaker array beamformer designs such as MECD and MSCD. Characterizing the projection sub-problem in MECD-PA and MSCD yielded a fast quadratic secular equation root-finding solution. The PA method accelerated convergence rates over the baseline DM, IPM, and SDP methods. GRPQ solutions exhibited more regular beam patterns than GRQ for a sample $3$-way system. Inequality constrained GDI can be considered in future works.

\appendix

\label{SEC:APPENDIX:LAGSOL}

The necessary conditions for $(\VEC{v}, \lambda)$ in \eqref{EQ:GRQ:CGRQ_HARD_PROJ} are given by
\begin{equation}
\begin{split}
\PR{\VEC{w} + \VEC{v}}^H \MAT{D} \PR{\VEC{w} + \VEC{v}} & = 0, 
\quad 
\VEC{v}  = \lambda \MAT{D}\PR{\VEC{w} + \VEC{v}}, 
\end{split}
\label{EQ:APPENDIX:LAGSOL:NEC_COND}
\end{equation}
where several useful relations are derived for characterizing the minimizer.
%%%%%%%
Expanding the equality constraint in \eqref{EQ:APPENDIX:LAGSOL:NEC_COND} relates the sum of symmetric terms to the sum of mixed terms:
\begin{equation}
\begin{split}
%%%%%%%
\VEC{w}^H \MAT{D} \VEC{w} + \VEC{v}^H \MAT{D} \VEC{v} & 
%%%%%%%%%%%5
 = - \PR{\VEC{v}^H \MAT{D} \VEC{w} + \VEC{w}^H \MAT{D} \VEC{v}}.
\end{split}
\label{EQ:APPENDIX:LAGSOL:CONSTR}
\end{equation}
%and multiplying $\PR{\VEC{w} + \VEC{v}}^H$ by $\VEC{v}$  and $\VEC{w}^H$ by $\VEC{v}$ in the gradient constraints gives
%\begin{equation}
%\begin{split}
%%%%%%%%
%\PR{\VEC{w} + \VEC{v}}^H \VEC{v}  = 0 \quad & \Rightarrow \quad \VEC{v}^H \VEC{v} = -\VEC{w}^H \VEC{v}, \\
%%%%%%%%%%
%\VEC{v}^H \VEC{v}  + \lambda \VEC{w}^H \MAT{D} \VEC{v} & = - \lambda \VEC{w}^H \MAT{D} \VEC{w},
%\end{split}
%\label{EQ:APPENDIX:LAGSOL:WV}
%\end{equation}
%%%%%%%%%
It is possible to relate the squared norm of $\VEC{v}$ from \eqref{EQ:APPENDIX:LAGSOL:NEC_COND} to the difference in squared norms under $\MAT{D}$ via \eqref{EQ:APPENDIX:LAGSOL:CONSTR}:
\begin{equation}
\begin{split}
\VEC{v}^H \VEC{v}  & = \frac{\lambda}{2} \PR{\VEC{v}^H \MAT{D} \VEC{w} + \VEC{w}^H \MAT{D} \VEC{v}} + \lambda \VEC{v}^H \MAT{D} \VEC{v} \\ 
%%%%%%%%%
& = \frac{\lambda}{2}  \PR{ \VEC{v}^H \MAT{D} \VEC{v}  - \VEC{w}^H \MAT{D} \VEC{w}}. 
%%%%%%%%%%
%\VEC{v}_1^H \VEC{v}_1  & =  \frac{\lambda_1}{2}  \PR{ \VEC{v}_1^H \MAT{D} \VEC{v}_1  - \VEC{w}^H \MAT{D} \VEC{w}}, \\
%%%%%%%%%%
%\VEC{v}_2^H \VEC{v}_2  & =  \frac{\lambda_2}{2}  \PR{ \VEC{v}_2^H \MAT{D} \VEC{v}_2  - \VEC{w}^H \MAT{D} \VEC{w}}.
\end{split}
\label{EQ:APPENDIX:LAGSOL:V1V1_V2V2}
\end{equation}
%%%%%%%
For $(\VEC{v}_1, \lambda_1)$, $(\VEC{v}_2, \lambda_2)$, the mixed products from \eqref{EQ:APPENDIX:LAGSOL:NEC_COND} are given:
%%%%%%%%
\begin{equation}
\begin{split}
\VEC{v}_2^H \VEC{v}_1 & =  \lambda_1 \VEC{v}_2^H  \MAT{D} \VEC{w} + \lambda_1 \VEC{v}_2^H \MAT{D} \VEC{v}_1, \\
%%%%%%%%
\VEC{v}_1^H \VEC{v}_2  & =  \lambda_2 \VEC{v}_1^H  \MAT{D} \VEC{w} + \lambda_2 \VEC{v}_1^H \MAT{D} \VEC{v}_2,
\end{split}
\label{EQ:APPENDIX:LAGSOL:MULT_EXP_1}
\end{equation}
%%%%
where their arithmetic mean is equivalent under conjugation $\VEC{v}_2^H \VEC{v}_1  = \frac{1}{2} \PR{\VEC{v}_2^H \VEC{v}_1  + \PR{\VEC{v}_1^H \VEC{v}_2 }^H}$ and $\VEC{v}_1^H \VEC{v}_2 = \PR{\VEC{v}_2^H \VEC{v}_1 }^H$:
\begin{equation}
\begin{split}
%%%%%%%%%%%%%
\VEC{v}_2^H \VEC{v}_1 & =  \frac{\lambda_1 \VEC{v}_2^H  \MAT{D} \VEC{w}  +  \lambda_2 \VEC{w}^H  \MAT{D} \VEC{v}_1}{2} + \frac{\lambda_1 + \lambda_2}{2}  \VEC{v}_2^H \MAT{D} \VEC{v}_1, \\
 %%%%%%%%%%%
 \VEC{v}_1^H \VEC{v}_2 & =  \frac{\lambda_1 \VEC{w}^H  \MAT{D} \VEC{v}_2  +  \lambda_2 \VEC{v}_1^H  \MAT{D} \VEC{w}}{2} + \frac{\lambda_1 + \lambda_2}{2}  \VEC{v}_1^H \MAT{D} \VEC{v}_2. \\
 %%%%%%%%%%%
\end{split}
\label{EQ:APPENDIX:LAGSOL:V2V1_MEAN}
\end{equation}
Summing \eqref{EQ:APPENDIX:LAGSOL:V2V1_MEAN} and substituting \eqref{EQ:APPENDIX:LAGSOL:CONSTR} gives
\begin{equation}
\begin{split}
%%%%%%%%%%%%%
\VEC{v}_2^H \VEC{v}_1 + \VEC{v}_1^H \VEC{v}_2   & = \frac{\lambda_1 + \lambda_2}{2} \PR{\VEC{v}_2^H \MAT{D} \VEC{v}_1 + \VEC{v}_1^H \MAT{D} \VEC{v}_2 - \VEC{w}^H \MAT{D} \VEC{w}}  \\
%%%%%%%
& - \frac{\lambda_1 \VEC{v}_2^H \MAT{D} \VEC{v}_2 + \lambda_2 \VEC{v}_1^H \MAT{D} \VEC{v}_1   }{2}.
 %%%%%%%%%%%
\end{split}
\raisetag{10pt}
\label{EQ:APPENDIX:LAGSOL:V2V1_V1V2_SUM}
\end{equation}
%%%%%%%%%
Subtracting \eqref{EQ:APPENDIX:LAGSOL:V2V1_V1V2_SUM} from the sum of $(\VEC{v}_1, \lambda_1)$, $(\VEC{v}_2, \lambda_2)$ substituted into \eqref{EQ:APPENDIX:LAGSOL:V1V1_V2V2} gives $\VEC{v}_1^H \VEC{v}_1 + \VEC{v}_2^H \VEC{v}_2 - \VEC{v}_2^H \VEC{v}_1 - \VEC{v}_1^H \VEC{v}_2 = \PR{\VEC{v}_1 - \VEC{v}_2}^H \PR{\VEC{v}_1 - \VEC{v}_2}$ which verifies \eqref{EQ:GRQ:CGRQ_HARD_PROJ_LAGRANGE_SECULAR_THM2}.
%%%%%%%%%%%
Next, observe that the arithmetic difference is equivalent under conjugation:
\begin{equation}
\begin{split}
%%%%%%%%%%%%%
0 & = \VEC{v}_2^H \VEC{v}_1 - \PR{\VEC{v}_1^H \VEC{v}_2}^H = \VEC{v}_1^H \VEC{v}_2 - \PR{\VEC{v}_2^H \VEC{v}_1}^H, \\
%%%%%%%
\Rightarrow 0 & = \frac{\lambda_1 \VEC{v}_2^H \MAT{D} \VEC{w} - \lambda_2 \VEC{w}^H \MAT{D} \VEC{v}_1 }{2} + \frac{\lambda_1 - \lambda_2}{2} \VEC{v}_2^H \MAT{D} \VEC{v}_1, \\
%%%%%%%%%%%
\Rightarrow 0 & = \frac{\lambda_1 \VEC{w}^H \MAT{D} \VEC{v}_2 - \lambda_2 \VEC{v}_1^H \MAT{D} \VEC{w} }{2} + \frac{\lambda_1 - \lambda_2}{2} \VEC{v}_1^H \MAT{D} \VEC{v}_2, \\
\end{split}
\label{EQ:APPENDIX:LAGSOL:V2V1_V1V2_DIFF}
\end{equation}
%%%%%%%%%
whereby taking the summation and substituting \eqref{EQ:APPENDIX:LAGSOL:CONSTR} gives
\begin{equation}
\begin{split}
0 & = \frac{\lambda_1 - \lambda_2}{2} \PR{ \VEC{v}_2^H \MAT{D} \VEC{v}_1 + \VEC{v}_1^H \MAT{D} \VEC{v}_2  - \VEC{w}^H \MAT{D} \VEC{w}} \\
%%%%%%%%%
& + \frac{\lambda_2 \VEC{v}_1 \MAT{D} \VEC{v}_1 -\lambda_1 \VEC{v}_2 \MAT{D} \VEC{v}_2 }{2}.
\end{split}
\label{EQ:APPENDIX:LAGSOL:V2V1_V1V2_DIFF_SUM}
\end{equation}
Subtracting \eqref{EQ:APPENDIX:LAGSOL:V2V1_V1V2_DIFF_SUM} from the difference  $\VEC{v}_1^H \VEC{v}_1 - \VEC{v}_2^H \VEC{v}_2$ after substitution into \eqref{EQ:APPENDIX:LAGSOL:V1V1_V2V2} verifies \eqref{EQ:GRQ:CGRQ_HARD_PROJ_LAGRANGE_SECULAR_THM1} after combining the terms.

%\newpage

%\bibliographystyle{IEEEbib}
\bibliographystyle{IEEEtran}
 
%\addtolength{\itemsep}{-0.05in}

%\vspace{-2.5mm}

\bibliography{IEEEabrv,refs}

\end{document}